\documentclass[12pt]{article}

\usepackage{epsfig}
\usepackage{amssymb}
\usepackage{amsmath}
\usepackage{color}
\usepackage{young}
\usepackage[vcentermath]{youngtab}

\makeatletter \@addtoreset{equation}{section} \makeatother

\newcounter{multieqs}

\newcommand{\be}{\begin{equation}}
\newcommand{\ee}{\end{equation}}
\newcommand{\bea}{\begin{eqnarray}}
\newcommand{\eea}{\end{eqnarray}}

\def\calD{{\cal D}}

\def\calZ{{\cal Z}}
\def\calW{{\cal W}}

\def\veps{\epsilon}

\begin{document}

\begin{flushright}\small
\texttt{HU-EP-13/77}\\
\texttt{DESY 2013-251}\\
\texttt{Hamburger Beitr\"age zur Mathematik 497}\\
\vspace{0.6cm}
\end{flushright}

\vspace{2pt}

\begin{center}

{\Large \bf Mayer-Cluster Expansion of Instanton Partition Functions and Thermodynamic Bethe Ansatz}
\vspace{20pt}

{\mbox {\large Carlo Meneghelli\,$^{a,b}$ 
 \ and Gang Yang\,$^{c}$}}\,%
\footnote{
{\tt  \ \tt \!\!\!carlo.meneghelli@desy.de,  gang.yang@physik.hu-berlin.de}
}


\begin{quote}
{\small \em
\begin{itemize}
\item[\ \ \ \ \ \ $^{a}$]
\begin{flushleft}
Fachbereich Mathematik, Universit\"{a}t Hamburg \\ 
Bundesstra$\beta$e 55, 20146 Hamburg
\end{flushleft}
\item[\ \ \ \ \ \ $^{b}$]
\begin{flushleft}
Theory Group, DESY \\ 
Notkestra$\beta$e 85, 22603 Hamburg, Germany
\end{flushleft}
\item[\ \ \ \ \ \ $^c$]
Institut f\"{u}r Physik, Humboldt-Universit\"{u}t zu Berlin\\
Newtonstra$\beta$e 15, 12489 Berlin, Germany

\end{itemize}
}
\end{quote}

\vspace{10pt}

\end{center}

\centerline{\bf { Abstract}}

\noindent In \cite{Nekrasov:2009rc} Nekrasov and Shatashvili pointed out that the ${\cal N}=2$ instanton partition function in a special limit of the $\Omega$-deformation parameters is characterized by certain 
thermodynamic Bethe ansatz (TBA) like equations. In this work we present an explicit derivation of this fact as well as generalizations to quiver gauge theories. 
To do so we combine various techniques like the iterated Mayer expansion, the method of expansion by regions, and the path integral tricks for non-perturbative summation.
The TBA equations derived entirely within gauge theory have been proposed to encode the spectrum of a large class of quantum integrable systems.
We hope that the derivation presented in this paper elucidates further this completely new point of view  on the origin, as well as on the structure, of TBA equations in integrable models.

\noindent

\newpage

\setlength{\parskip}{1pt}
\setcounter{tocdepth}{4}
\tableofcontents
\vspace{0.8cm}

\setcounter{tocdepth}{2}


\setlength{\parskip}{9pt}

\section{Introduction}

A number of remarkable connections have been observed between gauge theories and integrable systems.
They appear to be  useful to increase our understanding of both subjects.
On the one hand, using powerful integrability techniques one may hope to solve certain gauge theories non-perturbatively.
 On the other hand, gauge theory can help to formulate and solve integrable system.
A spectacular example is given by  planar ${\cal N}=4$ super Yang-Mills theory, for which in the last ten years or so tremendous progress has been achieved in solving the theory based  on the underlying integrability and on the AdS/CFT correspondence, see \cite{Beisert:2010jr} for a recent review. 
There exists another amazing connection between gauge theories and integrable models. 
The gauge theories in this case are  ${\cal N}=2$ supersymmetric gauge theories.
They do not need to be planar but the connection with integrable models is restricted to a special class of supersymmetric observables.
In this paper we focus on an important object in this class, the so-called instanton partition function and its relation with quantum integrable systems.

In the groundbreaking work of Seiberg and Witten \cite{Seiberg:1994rs, Seiberg:1994aj},  the exact solution for the  low energy effective action  of certain ${\cal N}=2$ gauge theories was proposed based on holomorphicity properties and electromagnetic duality. The low energy dynamics are encoded in a single object, called prepotential $\mathcal{F}(\vec{a})$.
 It is  a holomorphic function on the Coulomb moduli space, with coordinates $\vec{a}$,  and can be reconstructed from the so-called Seiberg-Witten (SW) curve and SW differential.
Shortly after, it was realized  that this description provides a direct connection between $\mathcal{N}=2$ gauge theories and classical algebraically integrable systems \cite{Gorsky:1995zq, Donagi:1995cf, {Itoyama:1995uj}}, see e.g.~\cite{D'Hoker:1999ft} for a pedagogical introduction.
These are essentially a complex analogue of integrable systems in the sense of Liouville.

The challenging program  of obtaining the Seiberg-Witten prepotential by a direct gauge theory calculation, developed on \cite{Moore:1997dj,Losev:1997bz, Moore:1998et},  was finalized in  \cite{Nekrasov:2002qd}.
This problem was solved using powerful localization techniques. Interestingly, this calculation produced a two-parameter, called $\epsilon_1$ and $\epsilon_2$, deformation of the prepotential.
The SW prepotential can  be obtained 
as
\be {\cal F}(\vec{a}; q) \,=\, \lim_{\epsilon_1, \epsilon_2 \rightarrow0} \epsilon_1 \epsilon_2 \log {\cal Z}(\vec a, \epsilon_1, \epsilon_2; q) \,, \ee
where we add the explicit dependence on the coupling constant $q$, but suppress dependence on further parameters such as masses.
The partition function $\mathcal{Z}$ receives tree level, one loop and instanton contributions. The latter part is usually referred to as Nekrasov instanton partition function.

The parameters $\epsilon_1,\epsilon_2$ correspond to a Lorentz rotation $\Omega$, thus the name $\Omega$-deformation, that encodes certain twisted boundary conditions for the four dimensional gauge theory. 
It was first introduced in \cite{Moore:1997dj,Losev:1997bz} in order to regularize the volume of the instanton moduli space.
The $\Omega$-deformation can be understood in a simple way by considering the five dimensional lift of the $\mathcal{N}=2$ theory, further compactified on a circle  \cite{Nekrasov:1996cz}.
In this set-up it is interpreted as  twisting of $\mathbb{R}^4$ by  a Lorentz rotation while going around the circle.
Introducing a two-parameter generalization of the prepotential triggered a huge progress.
An important example is the connection with topological strings. Upon  taking $\epsilon_1 = - \epsilon_2 = g_s$, the gauge theory partition function reproduces the topological string partition function with $g_s$ as genus parameter \cite{Iqbal:2003ix,Iqbal:2003zz}. The question of what is the topological string theory  analog of the $\epsilon_{1} \neq \epsilon_2$ case leads to the  definition of so-called ``refined" topological strings 
\cite{Hollowood:2003cv, Iqbal:2007ii}. Another spectacular example of progress driven by the calculation of the Nekrasov partition function is given by the Alday-Gaiotto-Tachikawa (AGT) correspondence \cite{Alday:2009aq}.

More recently, Nekrasov and Shatashvili \cite{Nekrasov:2009rc}  proposed   that upon taking the limit $\epsilon_2\rightarrow 0$ and
interpreting  $\epsilon_1$ as Plank constant, one obtains a correspondence between supersymmetric vacua
of a given gauge theory and eigenstates of the corresponding \emph{quantum} integrable model.
The relation between the SW prepotential and classical integrable systems is thus quantized.
This is usually called Nekrasov-Shatashvili (NS) limit and will be the main focus of this paper.
The central role in this correspondence is played by the so-called twisted superpotential ${\cal W}(\vec a, \epsilon_1;q)$ defined as
\be {\cal W}(\vec a, \epsilon_1; q) \,=\, \lim_{\epsilon_2 \rightarrow0} \epsilon_2 \log {\cal Z}(\vec a, \epsilon_1, \epsilon_2; q) \,. \label{defWINTRO} \ee
The Nekrasov-Shatashvili proposal is that, once this function is known, the eigenstates of  the quantum  integrable system are classified by solutions of the following quantization condition  
\be \exp \Big( {\partial {\cal W}(\vec a) \over \partial a_\lambda } \Big)= 1 \,, \qquad \qquad\lambda\,=\,1,\,\dots\,r\,, \ee
where $r$ is the rank of the gauge group.
These equations identify the twisted superpotential $\mathcal{W}$ with the so-called Yang-Yang (YY) function \cite{YangYang} of the quantum system. 
The proposed correspondence  provides an efficient general mechanism to define and solve quantum integral models.
Remarkably, it can be argued that the instanton part of the prepotential $\mathcal{W}_{\text{inst}}$, defined via  \eqref{defWINTRO}, can be characterized as the solution of certain non-linear integral equation 
of the  Thermodynamic Bethe Ansatz (TBA) type \cite{Zamolodchikov:1989cf}.
The main goal of this paper is to  develope some of the ideas presented in  \cite{Nekrasov:2009rc} to
give a more explicit derivation, as well as some generalizations,  of such TBA equations.

The proposal above originates as   some sort of extension of the so-called Bethe/gauge correspondence \cite{Nekrasov:2009uh,Nekrasov:2009ui}. 
The latter is based on the observation that the vacuum equations of  two dimensional $\mathcal{N}=4$
  gauge theories, broken to $\mathcal{N}=2$ by twisted masses,  coincides with Bethe equations for integrable models. 
The two dimensional twisted superpotential is equal to the YY function that encodes the Bethe equations and 
 Coulomb parameters correspond to the Bethe roots.
 The generators of chiral ring of the gauge theory \cite{Cecotti:1991me} are mapped to  Hamiltonians  of the integrable model, 
while their expectation values mapped to the corresponding eigenvalues.
In this way one obtaines a large  class of integrable models whose spectrum is characterized by traditional, possibly nested, Bethe equations.
Many integrable models do not belong to this class. The simplest example is given by the quantum Toda chain, see e.g.~\cite{Kozlowski:2010tv}, whose classical limit is connected to four dimensional  pure $SU(N)$ SYM. 
From the insight of the Bethe/gauge correspondence it is then natural to propose  \cite{Nekrasov:2009rc} the two dimensional twisted superpotential  which  corresponds to such integrable models.
It is the effective low energy action for the four dimensional   $\mathcal{N}=2$ gauge theory subject to an $\Omega$-background that preserves two dimensional $\mathcal{N}=2$ super-Poincar\'{e} symmetry, namely $\epsilon_2=0$.
This observation provides a stong motivation for the above correspondence.
An interpretation of this correspondence was given  using brane constructions in \cite{Nekrasov:2010ka}.
A further essential step in understanding the nature of the relation between quantum integrable systems and gauge theories has been presented in \cite{Nekrasov:2011bc}.

The proposal of Nekrasov and Shatashvili has inspired many other studies. Let us mention a few.
In \cite{Mironov:2009uv, Mironov:2009dv} it was shown that, similarly to the prepotential $\mathcal{F}$,
 the twisted superpotential ${\cal W}(\epsilon_1)$ can be obtained by calculating period integrals of a suitably deformed SW differential.
This analysis is also inspired by the  AGT correspondence, by which the NS limit corresponds to the semiclassical limit of  Liouville CFT \cite{Piatek:2011tp, Mironov:2012uh}.

In our work the Coulomb parameters will be assumed to be in generic positions. Extra considerations are needed if they take special values. For example in the conformal $SU(N)$ SYM with $N_f=2 N$, if the Coulomb parameters $a$
are set to be equal to  $m_{\text{fund}}-n\,\epsilon_1$, where $n\,\in\,\mathbb{Z}^{N}$, one can quantize the corresponding integrable system to obtain a  lenght $N$ spin-chain with infinite dimensional heighest weigth representations of $\mathfrak{sl}_2$  at each site \cite{Dorey:2011pa, Chen:2011sj, Chen:2012we}. Its spectrum is described in terms of traditional Bethe Ansatz equations.
Such developments triggered the discovery of a number of new dualities between various integrable models \cite{Bao:2011rc, Mironov:2012ba, Bulycheva:2012ct, Chen:2013jtk, Luo:2013nxa}.
The NS proposal has also inspired various  studies in (refined topological) string theories  \cite{Aganagic:2011mi, Huang:2011qx, Hellerman:2012zf, Huang:2012kn, Antoniadis:2013bja, Hatsuda:2013oxa, Kallen:2013qla,Bao:2013pwa} where the
the general $\Omega$-background plays a crucial role.

Despite the importance of this correspondence, 
the precise  mechanism by which  the instanton part of the twisted superpotential defined in \eqref{defWINTRO} turns out to be characterized as the solution of  TBA equations  is still to be elucidated.  In this paper, we will fill this gap.
In order to fully prove the NS's proposal, at least in some example, one should be able to show that the same TBA equation characterizes the spectrum of the corresponding integrable model.
In the case of pure $SU(N)$ SYM, which corresponds to the  periodic Toda chain with $N$ sites, this was achieved in \cite{Kozlowski:2010tv}. This interesting problem will be studied elsewhere \cite{US}.
In the following we  briefly outline the main ingredients used in our derivation of the TBA equations for \eqref{defWINTRO}, as well as the structure of the paper.

As pointed out in \cite{Nekrasov:2009rc}, it is convenient to start with the contour integral form of the instanton partition function.
In this representation the instanton partition  function can be interpreted as the partition function of a non-ideal gas of particles. The particular structure  of the two-particle interaction potential makes the study of the 
$\epsilon_2\rightarrow 0$ limit rather subtle. More precisely, this potential is the sum of a short-range (of order $\epsilon_2$) strongly attractive piece  and a long-range interaction part.
In Sections \ref{sec:simple} and \ref{originDILOG}, we  consider the simplified situation in which either the long- or short-range part is  set to zero. 
In order to study these simplified partition functions in the $\epsilon_2\rightarrow 0$ limit, we combine a number of techniques like Mayer expansion \cite{Mayer}  (a standard method in statistical mechanics)
 and the method of expansion by regions \cite{Beneke:1997zp} (a powerful method to compute Feynman integrals in small parameter expansions).
For the case with only long-range interactions, the $\epsilon_2\rightarrow 0 $ limit turns the logarithm of the partition function into a sum over certain tree graphs.
On the other hand, the free energy corresponding to only short-range interactions gives rise to  the dilogarithm function Li$_2$, which can be shown either by direct residue calculation of relevant integrals or via Mayer expansion together with the method of expansion by regions. 

In order to study the full partition function we find it convenient to use an \emph{iterated} version of Mayer expansion, see \cite{Gopfert:1981zu}. 
This expansion effectively creates a new partition function whose ``new particles"
 are \emph{clusters} of the original particles. The interaction within each cluster is governed by the short-range interaction, the one between different clusters by the long-range part.   
The iterated Mayer expansion thus produces an expression for the twisted superpotential $\mathcal{W}_{\text{inst}}$ as a  sum over tree graphs,  with vertices given by {\it clusters}. This expansion is carried over in some details in  Section \ref{sec:fullexpansion}.
The expression can be compared to high order in the instanton number with the expression coming from the solution to the TBA equation as discussed in Section \ref{perturbative_TBA}, providing direct non-trivial check of the equality.

There is an elegant way to prove that this equality holds to all orders in the instanton counting parameter $q$. It is based on rewriting the grand canonical partition function of the non-ideal gas in terms of a $(0+1)$-dimensional path integral. The analysis needs some special care as the potential has an unusual feature of depending in a singular way on $\hbar$, which is identified with $\epsilon_2$ in this case.
 A slight modification of the argument in  \cite{Polyakov}, together with the calculation of the contribution from the short range interactions corresponding to the dilogarithm, shows that the instanton partition function in the $\epsilon_2\rightarrow 0 $ limit is obtained by the saddle point evaluation of the  path integral. The saddle point equations are nothing but the TBA equations. This is explained in details in Section \ref{sec:nonpert}.

In Section \ref{sec:quiver} we present a generalization of the TBA equations corresponding to quiver gauge theories.
 More precisely, we consider  quivers characterized by a Dynkin diagram of $ADE$, or $\widehat{A}\widehat{D}\widehat{E}$  type. 
The twisted superpotential for such theories is shown to satisfy a set of coupled TBA equations with  one equation for each node of the  quiver and couplings corresponding to edges in the quiver.
The derivation is a simple extension of the one for the single gauge group case. This is so as the short range interaction, responsible for the clustering of particle, is non-vanishing only for particles corresponding to  the same gauge group factor in the quiver.

In order not to overload the main text, in Appendices we  include some review material together with  a few technical points concerning the derivation.
 A review of the contour integral form of the instanton partition function is given in Appendix \ref{app:contour_integral}. 
Some useful formulas are collected in Appendix \ref{App:UsefulFormula}. 
A discussion of the method of expansion by regions is given in Appendix \ref{app:region}.
 In Appendix \ref{App:regionsplusMayer}, we present an alternative derivation of a tree graphs expansion of the instanton partition function. 

The full partition function is a product of three terms $\mathcal{Z}=\mathcal{Z}_{\text{tree}}\,\mathcal{Z}_{\text{1-loop}}\,\mathcal{Z}_{\text{inst}}$.
In this paper we will be only concerned with the study of the instanton part ${\cal Z}_{\rm inst}$. For this reason from now on it will be simply denoted by ${\cal Z}$.

\subsection{NS's correspondence}
\label{sec:NS}

To complete the introduction, we present the integral representation of the instanton partition function 
for pure $SU(N)$  $\mathcal{N}=2$ super-Yang-Mills and the corresponding TBA equation.
The derivation of the TBA starting from the gauge theory expression of the instanton partition function is the main goal of the paper.

\subsubsection*{Instanton partition function}

The instanton partition function can be written  in a contour integral representation as
\be \label{Zinst} \calZ =  \sum_{k=0}^\infty \left({\epsilon \over \epsilon_1 \epsilon_2}\right)^k {q^k \over k!} \int  \prod_{I=1}^k {d \phi_I \over 2\pi i}  Q(\phi_I) \prod_{ J>I }^k \calD(\phi_{IJ}) \, ,  \ee
where $\phi_{IJ} = \phi_I - \phi_J$, $\veps = \veps_1 + \veps_2$, and
\be \calD(x) = {x^2 \over (x^2 - \epsilon_2^2)}{ (x^2 - \epsilon^2) \over  (x^2 - \epsilon_1^2)} \ ,
\label{defD} \ee
\be \label{QP} Q(x) = { 1 \over P(x) P(x+\epsilon) } \ , \qquad P(x) = \prod_{\lambda=1}^N (x-a_\lambda) \ . \ee
The origin of this expression is reviewed in Appendix \ref{app:contour_integral}.
The parameters entering this integrals, namely $\epsilon_{1,2}$ and $a_{\lambda}$ are taken to be real with a small positive imaginary part $i\, 0$.
The domain of integration above should be understood either as a real slice integration or equivalently, upon closing the integration in the upper-half plane, as a multiple contour integral.
In Appendix \ref{App:contourRES} we review how the residue evaluation of \eqref{Zinst} reproduces the representation of the instanton partition function as sum over $N$-tuples of Young diagrams.

We emphasize that the precise form of $Q(x)$ and $\calD(x)$ does not affect the derivation presented in this paper. This is the main reason why the generalization to quiver gauge theories is rather straightforward.
On the other hand the presence of the factor  ${x^2 \over x^2 - \epsilon_2^2}$ in ${\cal D}(x)$, which have a particularly singular limit for  $\epsilon_2$ small, will play a crucial role and 
will be responsible for the appearance of the  dilogarithm function in the TBA.

\subsubsection*{TBA form}

The claim of \cite{Nekrasov:2009rc} is that the twisted superpotential, defined as 
\be \calW = {\rm Limit}_{\veps_2 \rightarrow 0} \, ( \veps_2 \log \calZ) \,, \ee
can be written as the  critical value of the following integral functional:
\be {\cal Y}[\rho, \varphi] =  {1\over2} \int {d x \over 2\pi i}\,{d y \over 2\pi i} \, \rho(x) G(x-y) \rho(y) + \int  {dx \over 2\pi i} \Big[ \rho(x) \varphi(x) + {\rm Li}_2\left( q Q(x) e^{-\varphi(x)}\right) \Big]  \ , 
\label{YYdef} \ee
which is
\be \calW = {\rm Crit}_{\rho, \varphi}  \big[ {\cal Y}(\rho, \varphi) \big]  
= \int {dx \over 2\pi i} \Big[ -{1\over2} \varphi(x) \log\left(1 - q Q(x) e^{-\varphi(x)} \right)  + {\rm Li}_2\left( q Q(x) e^{-\varphi(x)}\right) \Big] \, , \label{tbaW} \ee
where $\varphi(x)$ is the solution of the following TBA-like equation
\be \varphi(x) = \int {d y \over 2\pi i} \, G(x-y) \log\left( 1-qQ(y) e^{-\varphi(y)} \right) . \label{tbaphi} \ee
$Q(x)$ is defined in (\ref{QP}), and the propagator $G(x)$ is related to $\calD(x)$ as
\be \label{defG} G(x) = {\rm Limit}_{\veps_2 \rightarrow0} {\calD(x) - 1 \over \veps_2} = {d \over dx} \log \left( {x+\epsilon_1 \over x- \epsilon_1} \right) \, . \ee
%

\section{Mayer-Cluster expansion}
\label{sec:Mayer-Cluster}

As mentioned in \cite{Nekrasov:2009rc}, the  contour integral form of the instanton partition function (\ref{Zinst})  can be interpreted (for each $k$) as the partition function of a one dimensional non-ideal gas of particles $\phi_1, \ldots, \phi_k$, subject to an external potential  and a pair-wise interaction potential respectively given by
\be \label{potentialU} U^{\rm ext}(x) = - \log\left( Q(x) {\epsilon \over \epsilon_1 \epsilon_2} \right) , 
\qquad \qquad 
V^{\rm int}(x) = - \log \left( \calD(x) \right) \, . \ee
Upon summing over the number of particles  $k$, the instanton partition function takes
 the form of a grand canonical partition function.
The free energy of this gas can be studied by Mayer expansion techniques \cite{Mayer} (see for example \cite{Brydges} for a nice introduction), as pointed out in \cite{Nekrasov:2009rc}. 
In this section, we will perform such kind of expansion in full details.
The limit of $\epsilon_2 \rightarrow 0$ appears to be rather subtle. 
In order to perform this limit we need to face the problem of  
studying the leading behavior of some multiple integral where a parameter is small.
It turns out that this can be conveniently studied by employing the method of expansion
by regions \cite{Beneke:1997zp} discussed in Appendix \ref{app:region}.
We introduce this method to provide a unified framework to study certain integrals,
but we stress that all the result of this section are also obtained
without employing such technique.

In order to analyze the behavior of  the  partition function (\ref{Zinst}) in the limit in which $\epsilon_2$ is small, it is convenient to split the function $ \calD(x)$, see \eqref{defD}, into two parts:  
\be \calD(x)\,=\,\frac{x^2}{x^2-\epsilon_2^2}\, \widetilde{\calD}(x)\,,\qquad \qquad
\widetilde{\calD}(x)\,=\,\frac{x^{2}-\epsilon^2}{x^{2}-\epsilon_1^2}\,.
\label{splittingPOT} \ee
The reason of such decomposition is as follows. 
The factor $\frac{x^2}{x^2-\epsilon_2^2}$ corresponds to a pair-wise interaction which is strong and attractive at distances of order $\epsilon_2$ and rapidly decreases at large distances.
The remaining factor $\widetilde{\calD}(x)$  corresponds to a pair-wise interaction which is different from zero only at distances of order $\epsilon_1$.
Thus  \eqref{splittingPOT} corresponds to splitting the potential into short- and long-range parts.
The natural candidate to study such kind of potentials is the so-called {\it iterated} Mayer expansion \cite{Gopfert:1981zu}. 
As we will see in  Section \ref{sec:fullexpansion}, this effectively creates a new grand canonical partition function whose ``new particles'' correspond to  clusters of the original particles.  
We will start by considering some simplified situation.

\subsection{Only long range interactions}
\label{sec:simple}

\begin{center}
\begin{figure}[t]
\centerline{\includegraphics[height=2.6cm]{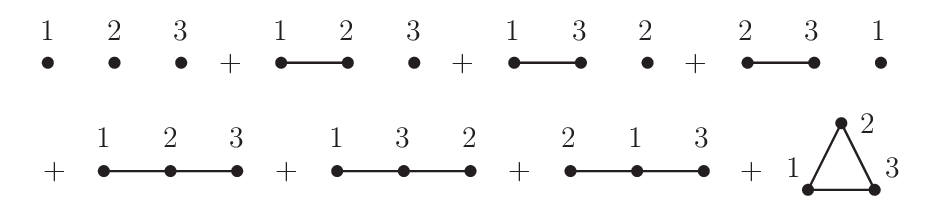} } 
\caption{\it The graph expansion for the $k=3$ case. The second line contains all connected graphs.} 
\label{fIJ_k=3}
\end{figure}
\end{center}

Let us first consider a simplified version of  (\ref{Zinst}) without the factors\footnote{We also set $\frac{\epsilon}{\epsilon_1}$ to one.} ${\phi_{IJ}^2 \over \phi_{IJ}^2 - \epsilon_2^2}$, i.e
\be {\cal Z}_{\rm Long} \,:=\, \sum_{k=0}^\infty\,  {q^k \over k!}\,
 \int   \prod_{I=1}^k 
{d \phi_I \over 2\pi i}\,{Q(\phi_I) \over \epsilon_2} \, \prod_{1\,\leq\,I < J\leq\,k}\,\widetilde{\calD}(\phi_{IJ})  \,.
\label{Zsimple} \ee
The basic idea of Mayer expansion, see e.g.~\cite{Brydges},  is to introduce the function $\widetilde{f}_{IJ}$ as
\be \label{FG} \widetilde{\calD}(\phi_{IJ}) \,= \,1 + \widetilde{f}_{IJ}\,, \ee
and  expand the interaction products as
\begin{equation}
  \prod_{1\,\leq\,I < J\leq\,k}\left(1+\widetilde{f}_{IJ}\right)\,=\,1\,+\,\sum_{I<J}\, \widetilde{f}_{IJ}\,+\,\sum_{I<J,I'<J'}\,\widetilde{f}_{IJ}\widetilde{f}_{I'J'}\,+\dots\,. \label{expansionD}
\end{equation}
Each monomial in the right hand side of this equation can be visualized as a graph (not necessarily connected) on the set $[k]=\{1,2,\dots,k\}$. 
More precisely, each particle in  $\{1,2,\dots,k\}$ corresponds to a vertex, and to each factor $\widetilde{f}_{IJ}$  we associate an edge between particle $I$ and $J$  in the corresponding graph. More explicitly,
\be \widetilde f_{IJ} \ \  :=   \begin{tabular}{c}{\includegraphics[height=.8cm]{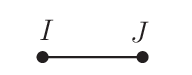}}\end{tabular}  \ee
As an example for the $k=3$ case, we have the expansion in terms of graphs shown in Figure \ref{fIJ_k=3}. It is clear from the left hand side of \eqref{expansionD} that there are no multiple edges between two vertices, or edges connecting one vertex to itself. 

%

\begin{center}
\begin{figure}[h]
\centerline{\includegraphics[height=4.cm]{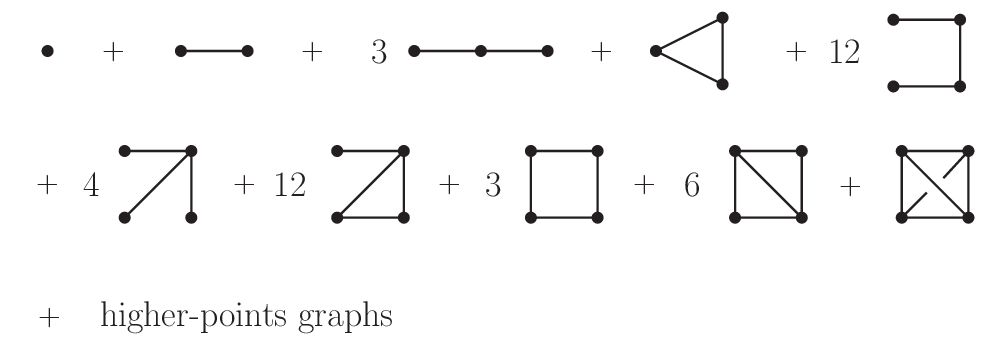} } 
\caption{\it
Connected graphs in the Mayer expansion. 
The coefficients correspond to how many different graphs of the given topology belong to ${\cal G}_c^{[k]}$. They are obtained from one another by  relabeling the vertices.}
\label{connect_graph}
\end{figure}
\end{center}

This expansion is particularly useful, as the logarithm of the grand canonical partition function can be formally given as a sum over {\it connected}  graphs \cite{Mayer} (see for example Appendix A of \cite{Brydges} for a simple derivation):
\be \label{simple} \log {\cal Z}_{\text{Long}}\,=\,
\sum_{k=1}^\infty \,{q^k \over k!}\, \int\, \prod_{I=1}^k\,  {d\phi_I \over 2\pi i}  {Q(\phi_I) \over \epsilon_2}\,
 \sum_{g\,\in \,{\cal G}_c^{[k]}}\, \prod _{e(I,J)\,\in\, g} \widetilde{f}_{IJ}   \ . \ee
Here ${\cal G}_c^{[k]}$ denotes the collection of connected graphs on the set $[k]$  and  $e(I,J)$ belongs to the  graph $g$ if there is an edge between  the vertices $I$ and $J$.  
 Graphs up to four points are shown in Figure \ref{connect_graph}.
 
%

\begin{center}
\begin{figure}[h]
\centerline{\includegraphics[height=2.7cm]{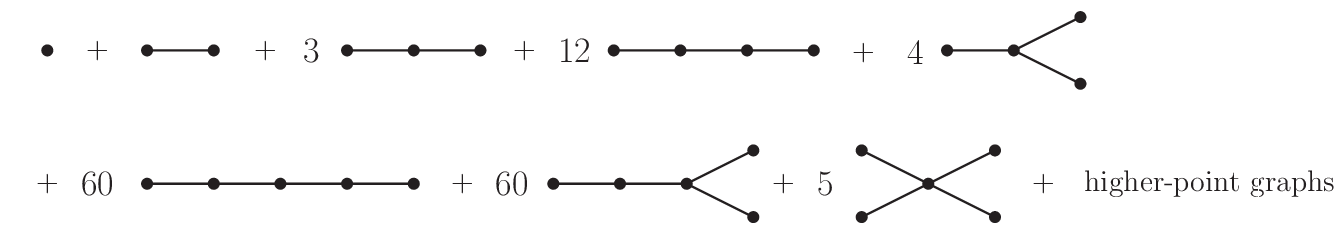} } 
\caption{\it
Connected tree graphs up to five-point. The coefficients corresponds to how many different graphs of the given topology belong to ${\cal T}_c^{[k]}$, which are related to each other by  relabeling the vertices.} 
\label{connect_tree_graph}
\end{figure}
\end{center}

We stress that  \eqref{simple} is an exact relation as a formal power series\footnote{The interesting question of convergence of the Mayer expansion can be addressed in various ways, see e.g. \cite{Brydges}. 
Here, we will not consider this problem or the analog convergence issue for the solution to the TBA.} in $q$.
 In the limit of small $\epsilon_2$ one has 
\be  \widetilde{f}_{IJ}\, =\,  \epsilon_2 \,G(\phi_{IJ}) + {\cal O}(\epsilon_2^2) \, , \label{FG2}\ee
where $G$ is defined in (\ref{defG}). 
As each factor of $\widetilde{f}$ contributes one power of $\epsilon_2$,
 the leading contribution to the sum in \eqref{simple} is given only by connected {\it tree} graphs $\mathcal{T}^{[k]}_c$.
 Such graphs have the minimal number of edges ($k-1$) among the connected graphs. The sum of these tree graphs can be shown to be convergent. Collecting the powers of $\epsilon_2$ we conclude that 
\be \label{simpleLIMIT} \log {\cal Z}_{\rm Long}\,=\,\frac{1}{\epsilon_2}
\,\sum_{k=1}^\infty \,{q^k \over k!}\, \int\, \prod_{I=1}^k\,  {d\phi_I \over 2\pi i}  Q(\phi_I)\,
 \sum_{g\,\in \,{\cal T}_c^{[k]}}\, \prod _{e(I,J)\,\in\, g} \widetilde{f}_{IJ} \,+\,  {\cal O}(\epsilon_2^0)  \,. \ee
Graphs up to five points are shown in Figure \ref{connect_tree_graph}.
Notice that the leading behaviour of the free energy \eqref{simpleLIMIT} is at order $\frac{1}{\epsilon_2}$.
This fact may be not obvious from the definition \eqref{Zsimple} where for each $k$ the leading contribution is proportional to $\frac{1}{\epsilon_2^k}$.
We will see that the similar behaviour applies to the more complicated situations analized in the following.

%
%
\subsection{Only  short range interactions \label{sec:Li2}}
\label{originDILOG}

Next we study another simplified version of (\ref{Zinst}).
Namely, we  set $\widetilde{\mathcal{D}}$, defined in \eqref{splittingPOT}, to zero and consider
\be \calZ_{\text{Short}}\, :=\,
  \sum_{k=0}^\infty \,{q^k \over k!} \,\int   \,  \prod_{I=1}^k {d \phi_I \over 2\pi i}\,{ Q(\phi_I) \over \epsilon_2} \,
\prod_{1\,\leq\,I < J\,\leq k} {\phi_{IJ}^2 \over \phi_{IJ}^2 - \epsilon_2^2} \,.
\label{ZwithF=0} \ee
The main achievement of this subsection is to show that in the limit of small $\epsilon_2$ the logarithm of \eqref{ZwithF=0} is given by
\be \log\calZ_{\text{Short}}\, = \,\frac{1}{\epsilon_2}\, \int {d\phi_0 \over 2 \pi i} \, {\rm Li}_2[qQ(\phi_0)]  \,+\,{\cal O}(\epsilon_2^0) \, . 
\label{resultLi2}\ee
 We will see that, as opposed to the case of long range interactions considered in the previous subsection,
 the right hand side of \eqref{resultLi2} is not given by summing over tree graphs but  by a single local term.
As will be explained in details in Section \ref{sec:fullexpansion}, this is the underlying mechanism by which the 
short-range interactions turn a group of particles  into a single effective particle.
The result \eqref{resultLi2} explains why the dilogarithm function appears in the TBA. 
 It will also be  an essential input in the all order proof presented in Section \ref{sec:nonpert}. 
Considering its importance, we will prove \eqref{resultLi2} in two different ways.

\subsubsection{Dilogarithm from a sum over residues}

In this subsection we will prove \eqref{resultLi2}  by direct evaluating  \eqref{ZwithF=0} and \eqref{resultLi2}
as a sum over residues. We find it instructive to first consider the partition function for the $U(1)$ gauge theory, i.e.~we
evaluate \eqref{ZwithF=0} for 
\begin{equation}
 Q(x)\,=\,\frac{1}{(x-a-i\,0)(x-a+\epsilon+i\,0)}\,,
\label{QN=1}
\end{equation}
compare to \eqref{QP}.
In \eqref{QN=1} $a$  and $\epsilon$ are real and we wrote explicitely the $i\,0$ prescription.
For each $k$ in \eqref{ZwithF=0}, there is only one residue (up to permutation of the integration variables) in the upper half plane,
compare to the general discussion in Appendix \ref{App:contourRES}.
It is given by $(\phi_1,\dots,\phi_k)=(a,a+\epsilon_2,\dots,a+(k-1)\epsilon_2)$. It follows that
\begin{equation}
 \mathcal{Z}^{U(1)}_{\text{Short}}
\,=\,\sum_{k=0}^{\infty}\,\frac{q^k}{k!}\,\frac{1}{\epsilon^k_2}\,\prod_{I=1}^k\,
\frac{1}{\epsilon_1+I\,\epsilon_2}\,.
\label{ZU1fromRES}
\end{equation}
One can also directly calculate the one dimensional integral 
\begin{equation}
 \int\,\frac{d \phi_0}{2\pi i}\,Q(\phi_0)^\ell\,=\,\frac{(-1)^{\ell+1}}{\epsilon^{2\ell-1}}\,\binom{2\ell-2}{\ell-1}\,,
\label{QtothenU1}
\end{equation}
where $Q(x)$ is given in \eqref{QN=1}. 
Using the identity \eqref{formulaforres} one concludes the the logarithm of \eqref{ZU1fromRES} in the $\epsilon_2\rightarrow 0$
limit is indeed given by \eqref{resultLi2} with \eqref{QtothenU1}.

For the more general $U(N)$ theory, it  is easy to classify the poles contributing to  \eqref{ZwithF=0} following 
the same  reasoning as in Appendix \ref{App:contourRES}. For fixed $k$, residues are classified, up to permutation of the $k$
particles, by  a set of integers  $\{ s_1, s_2, \ldots, s_N \}$ such that $\sum_\lambda s_\lambda = k$.
The corresponding pole is given by
\begin{equation}
\phi_{\kappa_{\lambda}+I_{\lambda}}\,=\,a_{\lambda}+(I_{\lambda}-1)\epsilon_2\,,
\qquad I_{\lambda}=1,\dots,s_{\lambda}\,,\qquad 
\kappa_{\lambda}=\sum_{\mu=1}^{\lambda-1}\,s_{\mu}\,,
\label{poleF=0}
\end{equation}
and  $\lambda=1,\dots,N\,$. As opposed to the full partition function (\ref{Zinst}), 
for which residues are classified by $N$-tuple of Young tableaux with a total number of boxes equal to $k$,
in the simplified integrals  \eqref{ZwithF=0} only Young tableaux with one row, whose length is denoted by $s_{\lambda}$,
contribute. It is straigthforward to calculate and collect all residues, see Appendix \ref{App:UsefulFormula}, to obtain
\begin{equation}
  \mathcal{Z}_{\text{Short}}\,=\,
\sum_{s_1,\dots,s_{N}\geq 1}\,
\prod_{\lambda=1}^N\,
\left(
\frac{q^{s_{\lambda}}\,\mathfrak{R}^{(\lambda)}_{s_\lambda}}{s_{\lambda}!\,\epsilon_2^{s_{\lambda}}}\right)\,
\prod_{\lambda < \lambda'}\prod_{I_\lambda, I_{\lambda'}} {(a_{\lambda \lambda'} + (I_\lambda - I_{\lambda'}) \epsilon_2)^2 \over (a_{\lambda \lambda'} + (I_\lambda - I_{\lambda'}) \epsilon_2)^2 - \epsilon_2^2}\,,
\label{ZD=0res}
\end{equation}
 where
\begin{equation}
\mathfrak{R}^{(\lambda)}_s\,:=\,
\prod_{I=1}^s\,\frac{Q^{(\lambda)}(a_{\lambda_1}+(I-1)\epsilon_2)}{\epsilon_1+I\epsilon_2}\,,\qquad
 Q^{(\lambda)}(x)\,=\,\frac{1}{P^{(\lambda)}(x)P^{(\lambda)}(x+\epsilon)}\,,
\label{ZD=0res2}
\end{equation}
and $P^{(\lambda)}(x)\,=\,\prod_{\mu\neq\lambda}(x-a_{\mu})$.
The structure of the result \eqref{ZD=0res}-\eqref{ZD=0res2} represents a simple generalization of the $U(1)$ 
computation \eqref{ZU1fromRES}.
The $\epsilon_2\rightarrow 0$ limit of the  logarithm of \eqref{ZD=0res} can be readily obtained using the relation \eqref{usufulresidue2}. One recognizes that
\begin{equation}
 \log\,\mathcal{Z}_{\text{Short}}\,=\,\frac{1}{\epsilon_2}\,
\sum_{\lambda=1}^N\,\sum_{\ell=1}^{\infty}\,\frac{q^{\ell}}{\ell^2}\,\text{Res}_{\phi_0=a_{\lambda}}\left[Q^\ell(\phi_0)\right]\,+{\cal O}(\epsilon_2^0)\,\,.
\end{equation}
This result coincides with the evaluation of \eqref{resultLi2} by residues.
 This calculation provides a direct proof of  \eqref{resultLi2}.

\subsubsection{By Mayer expansion and separation of regions}
\label{byregions}

We now calculate (\ref{ZwithF=0}) for small $\epsilon_2$ by first applying  Mayer expansion  and then  the so-called 
method of expansion by regions. 
The first step is to decompose the interaction factor as
\be {\phi_{IJ}^2 \over \phi_{IJ}^2 - \epsilon_2^2} = 1 + f_{IJ} \,, \qquad f_{IJ} := {\epsilon_2^2 \over \phi_{IJ}^2 - \epsilon_2^2} \, . \label{f_IJ} \ee
In a similar way as \eqref{simple}, the logarithm of (\ref{ZwithF=0}) is then given by
\be \log {\cal Z}_{\text{Short}}
=  \sum_{k=1}^\infty {q^k \over k! \epsilon_2^k} \int  \prod_{I=1}^k {d \phi_I \over 2\pi i} \, {Q(\phi_I) } \sum_{g\,\in\, {\cal G}_c^{[k]}}\, \prod _{e(I,J)\in g} f_{IJ}   \, . 
\label{LogZF=0}
\ee

To show that this gives \eqref{resultLi2}, 
we need to evaluate the integrals entering \eqref{LogZF=0} as a  Laurant series in $\epsilon_2$. 
We are actually interested only in the terms with leading negative powers of $\epsilon_2$.
We use a powerful method, usually applied to the evaluation of Feynman integrals,  called \emph{expansion by regions} introduced in \cite{Beneke:1997zp}. 
The method goes as follows: $(1)$ divide the integration domain into regions and expand the integrand in a Taylor series in small parameters, 
 $(2)$ extend the integration to the full domain of integration, $(3)$ set to zero scaleless integrals.
 This method turns out to be particularly efficient to calculate the leading term for the integrals  \eqref{LogZF=0}. Concerning these contributions, we do not need to apply the somewhat subtle point $(3)$ above. A more detailed discussion of this method will be given in Appendix \ref{app:region}.

\begin{center}
\begin{figure}[t]
\centerline{\includegraphics[height=1.1cm]{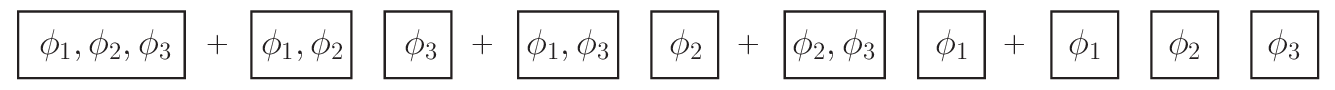} } 
\caption{\it
The decomposition of the contribution for $k=3$ case. Each square represents a cluster labeled by corresponding particles in the cluster.} 
\label{cluster_3}
\end{figure}
\end{center}

To identify the set of relevant regions, 
we note that the  Taylor expansion of the interaction term ${\epsilon_2^2 \over \phi_{IJ}^2 - \epsilon_2^2}$ starts 
 at  order  $\epsilon_2^2$ except for the  \emph{region} in which $\phi_{IJ}$ is of order $\epsilon_2$.
It is then natural to expect that regions are classified as follows. 
Let $\mathcal{B}^{[k]}$ denotes the set of grouping of $k$  labeled particles in clusters, see figure \ref{cluster_3} for the $k=3$ case. For each grouping  we define the corresponding region as
\begin{equation}
\begin{cases}
 |\phi_{IJ}|\, \sim\,\epsilon_2 & \text{if $I,J$ are in the same cluster,}\\
 |\phi_{IJ}|\, \gg\,\epsilon_2 & \text{if $I,J$ are in different clusters.}
\end{cases}\,
\end{equation}
%

%
%

The next step is to Taylor expand the integrand in each region. We denote by $\mathbb{T}^{(\mathsf{b})}\left[\dots\right]$ the operation of Taylor expaning $\dots$ in the region corresponding to the grouping  $\mathsf{b}\,\in\,\mathcal{B}^{[k]}$.  We have
\begin{equation}
\mathbb{T}^{(\mathsf{b})}\left[ f_{IJ}\right]\,=\,
\begin{cases}
 {\cal O}(1)  & \text{if $I,J$ are in the same cluster,}\\
{\cal O}(\epsilon_2^2) & \text{if $I,J$ are in different clusters,}
\end{cases}\,
\label{Tbf}
\end{equation}
and
\begin{equation}
 \mathbb{T}^{(\mathsf{b})}\left[ \prod_{I=1}^k Q(\phi_I)\right]\,=\,
 \prod_{\mathbf{Y}\,\in\,\mathsf{b}}\,Q(\phi_{\mathbf{Y}})^{|\mathbf{Y}|} \,+\,\dots\,\qquad 
\phi_{\mathbf{Y}}\,:=\,\frac{1}{|\mathbf{Y}|}\,\sum_{I\,\in\,\mathbf{Y}}\,\phi_I\,,
\label{TbQQ}
\end{equation}
the product on the right hand side runs over the clusters in the grouping $\mathsf{b}$ and $|\mathbf{Y}|$ denotes the number of particles in the cluster $\mathbf{Y}$.
From  \eqref{Tbf}-\eqref{TbQQ} it is clear that, for each $k$, the leading contribution to  \eqref{LogZF=0} comes from the region in which all $\phi_I$ are in the same cluster. 
Indeed, if there were two or more clusters, the  $f_{IJ}$ factor, with $I,J$ in different clusters, would decrease the power of $\epsilon_2$, compare to \eqref{Tbf}.
An explicit example of this  expansion of $k=3$  is given in appendix \ref{app:region}.
As we will shortly see, the leading contribution to \eqref{LogZF=0} starts at order  $1/\epsilon_2$ for each $k$. This fact is not obvious from the form  \eqref{LogZF=0}.
 
By further separating the integration over the ``center of cluster'' coordinate $\bar{x}$ as
\begin{equation}
 \int\,\prod_{I=1}^sdx_I\,F(x_1,\dots,x_s)\,=\,
\int\,d\bar{x}\, \int\,\prod_{I=1}^s dx'_I \,\delta\left(\frac{1}{k}\sum_{I=1}^s\,x'_I\right)\,
F(\bar{x}+x'_1,\dots,\bar{x}+x'_s)\,,
\label{introduceCenter}
\end{equation}
where we applied the change of variables 
\be \bar x := {1\over k} \sum_{I=1}^s x_s \,, \qquad x'_I := x_I - \bar x \,, \ee
one can rewrite \eqref{LogZF=0} as
\be 
\log {\cal Z}_{\text{Short}}
\,=\,\frac{1}{\epsilon_2} \sum_{k=1}^\infty {q^k \over k!}\,\int  d \phi_0 \,Q^k(\phi_0)\,\mathcal{J}_k\,+\,{\cal O}(\epsilon_2^0)\,
\label{logZregionsLi2}
\ee
where
\be
\mathcal{J}_k\,:=\,
 \,{1\over \epsilon^{k-1}_2} \int  \prod_{I=1}^k {d \phi_I \over 2\pi i} \,\delta\left(\frac{1}{k}\sum_{I=1}^k\,\phi_I\right) \sum_{g\,\in\, {\cal G}_c^{[k]}}\, \prod _{e(I,J)\in g} f_{IJ}   \, .
\label{defJk}
\ee
Notice that in \eqref{defJk} all connected graphs, independently on the number of edges, contribute to the leading term.
The next observation to be made is that 
\be
 \text{if $g$ is  not connected}\qquad
\,{1\over \epsilon^{k-1}_2} \int  \prod_{I=1}^k {d \phi_I \over 2\pi i} \,\delta\left(\frac{1}{k}\sum_{I=1}^k\,\phi_I\right) \prod _{e(I,J)\in g} f_{IJ}\,=\,0   \, . 
\label{zeronotconnected}
\ee
The validity of this statement can be easily argued as follows.
 For each connected component in $g$ we can define its center as the average of the $\{\phi_I\}$ in that connected component.  The integrand in \eqref{zeronotconnected} does not depend on the distance between the centers.
As the integral is calculated by residue, it trivially vanishes in this case as, after integrating trivially the delta function, the $(k-1)$-dimensional residue is absent.
The identity \eqref{zeronotconnected} implies that  in \eqref{defJk}, we can replace the sum over connected graphs with the sum over all graphs.
Finally using the relation  
\be
\sum_{g\,\in\, {\cal G}^{[k]}}\, \prod _{e(I,J)\in g} f_{IJ}\,=\,\prod_{1\leq I< J\leq k}(1+f_{IJ})\,,
\ee
where ${\cal G}^{[k]}$ is the set of all graphs on $[k]$, 
we conclude that $\mathcal{J}_k=\mathcal{I}_k$, where, using \eqref{f_IJ},
\begin{equation}
\mathcal{I}_k\,:=\,
 \,{1\over \epsilon^{k-1}_2} \int  \prod_{I=1}^k {d \phi_I \over 2\pi i} \,\delta\left(\frac{1}{k}\sum_{I=1}^k\,\phi_I\right)\, \prod_{1\leq I< J\leq k} {\phi_{IJ}^2 \over \phi_{IJ}^2 - \epsilon_2^2} \,=\,\frac{1}{2 \pi i }\,\frac{k!}{k^2}\,.
\label{resultJk} 
\end{equation}
Notice that the evaluation of this integral is exact, see \cite{Moore:1998et} for a derivation.
Using this result we recognize that \eqref{logZregionsLi2} is equal to 
\be 
\log {\cal Z}_{\text{Short}}
\,=\,\frac{1}{\epsilon_2}\, \sum_{k=1}^\infty {q^k \over k^2}\,\int  \frac{d \phi_0}{2 \pi i} \,Q^k(\phi_0)\,+\,{\cal O}(\epsilon_2^0)\,\, = \,\frac{1}{\epsilon_2}\, \int {d\phi_0 \over 2 \pi i} \, {\rm Li}_2[qQ(\phi_0)]  \,+\,{\cal O}(\epsilon_2^0) \,.
\ee
 
\subsection{Expansion of the instanton partition function}
\label{sec:fullexpansion}

We are now ready to consider the full instanton partition function  (\ref{Zinst}). 
We will combine the considerations from the previous sections. In Section \ref{sec:simple} we learned  that the $\epsilon_2\rightarrow 0$ limit singles out certain tree level graphs.
In  Section \ref{originDILOG} we learned that the factors $\frac{\phi_{IJ}^2}{\phi_{IJ}^2-\epsilon_2^2}$ produce the effect of combining particles into clusters.
To exploit the combination of these two mechanisms in the most transparent way we find it convenient to use the so-called iterated Mayer expansion advocated at the beginning of our analysis.
We start by reviewing this expansion. Based on this iterative expansion, by combining the discussion from the two previous sections, one obtains a tree graph expansion for the full instanton pationtion as given in Section \ref{sec:ourcase}. 
In Appendix \ref{App:regionsplusMayer} we show that the same result can be obtained by first applying  the method of expansion by  regions to the original partition function and then exploiting some combinatorics to conclude that  only certain connected graphs contribute to the free energy. 

\subsubsection{Iterated Mayer expansion}
\label{iteratedMayer}

We start with a review of the iterated Mayer expansion \cite{Gopfert:1981zu}. Consider the partition function
\begin{equation}
 \mathcal{Z}\,=\,\sum_{k=0}^{\infty}\,\frac{q^k}{k!}\int
\,\prod_{I=1}^k d\phi_I A(\phi_I)\prod_{1\leq I<J\leq k}\left(1+\mathcal{F}_{IJ}\right)\,,\qquad
1+\mathcal{F}_{IJ}\,=\,e^{-\left(V^{\text{int}}_S(\phi_{IJ})+V^{\text{int}}_L(\phi_{IJ})\right)}\,,
\label{Zforiterated}
\end{equation}
where, as in \eqref{splittingPOT}, we split the pair-wise interaction  potential in  a short and long range part.
Next we  introduce
\begin{equation}
\mathfrak{g}_{\mathbf{Y}}\,=\,
\prod_{\substack{I,J\,\in\,\mathbf{Y}\\I<J}}\,\left(1+\,e^{-V^{\text{int}}_L(\phi_{IJ})}\right)\,,\,\qquad 
 1+f_{IJ}\,=\,e^{-V^{\text{int}}_S(\phi_{IJ})}\,,
\end{equation}
\begin{equation}
 1+\mathfrak{f}_{\mathbf{Y}_a,\mathbf{Y}_b}\,=\,\prod_{I_a\,\in\,\mathbf{Y}_a}\prod_{I_b\,\in\,\mathbf{Y}_b}
\,e^{-V^{\text{int}}_L(\phi_{I_aI_b})}\,,\qquad a\neq b
\end{equation}
where $\mathbf{Y}$ denotes a set of particles or, in other words, a cluster.
Iterated Mayer expansion is the statement that the free enery can be expanded as
\begin{equation}
\log\, \mathcal{Z}\,=\,\sum_{k=1}^{\infty}\,\frac{q^k}{k!}\,
\sum_{\ell\geq 0}\,\sum_{\{\mathbf{Y}_1,\dots,\mathbf{Y}_\ell\}\,\in\,\mathcal{B}^{[k]}_{\ell}}\,
\int\,\prod_{I=1}^k d\phi_I \,\,
\mathfrak{S}_{\{\mathbf{Y}_1,\dots,\mathbf{Y}_{\ell}\}}\,,
\label{Mayer}
\end{equation}
where
\begin{equation}
\mathfrak{S}_{\{\mathbf{Y}_1,\dots,\mathbf{Y}_{\ell}\}}\,=\,
\left(\prod_{a=1}^{\ell}\,\mathfrak{g}_{\mathbf{Y}_a}\,S_{\mathbf{Y}_a}\right)\,
\sum_{g\,\in\,\mathcal{G}^{[\ell]}_c}
\,\prod_{e(a,b)\,\in\, g}\,\mathfrak{f}_{\mathbf{Y}_a,\mathbf{Y}_b}\,,
\label{Sgothicdef}
\end{equation}
\begin{equation}
S_{\mathbf{Y}}\,=\,\prod_{I=1}^k A(\phi_I)\,
\sum_{g\,\in\,\mathcal{G}^{\mathbf{Y}}_c}\,
\prod_{e(I,J)\,\in\, g}\,f_{IJ}\,,
\label{Sdef}
\end{equation}
The sum in \eqref{Mayer} is taken over $\mathcal{B}^{[k]}_{\ell}$:  groupings of $k$ (labeled) particles into $\ell$ clusters. As the notation may need some time to be digested, in Appendix \ref{App:ExampleIteratedMayer} we spell out the  definitions for $k=2,3$. Note that although the cluster here is in a different context, the picture of grouping is similar to that used before in expansion by regions, see for example figure \ref{cluster_3} for $k=3$ case.

\subsubsection{Iterated Mayer expansion for the Nekrasov partition function}
\label{sec:ourcase}

We can apply the iterated Mayer expansion reviewed in Section \ref{iteratedMayer}
to the full partition function, compare  \eqref{Zforiterated} to \eqref{Zinst}, \eqref{splittingPOT}.
Once this is done we need to evaluate the leading contribution for small $\epsilon_2$ to 
 integrals of the type \eqref{Mayer}.
The crucial  observation is that the   Taylor expansion $\tilde{D}(\phi_{IJ})-1\sim \,\epsilon_2\,G(\phi_{IJ})$
 is valid everywhere in the domain of integration\footnote{ 
In the language of the method of expansion by regions we would say that the leading term in the Taylor expansion of $\tilde{D}(\phi_{IJ})-1$ is the same in  \emph{any region}.}.
This immidiately implies that in the limit of small $\epsilon_2$ one can write, see definition \eqref{Sgothicdef},
\begin{equation}
\mathfrak{S}_{\{\mathbf{Y}_1,\dots,\mathbf{Y}_{\ell}\}}\,=\,\epsilon_2^{\ell-1}\,
\prod_{a=1}^{\ell}\,S_{\mathbf{Y}_a}\,
\sum_{g\,\in\,\mathcal{T}^{[\ell]}_c}
\,\prod_{e(a,b)\,\in\, g}\,\prod_{\substack{I_a\,\in\,\mathbf{Y}_a\\I_b\,\in\,\mathbf{Y}_b}}
G(\phi_{I_aI_b})+\dots
\label{equationTREE}
\end{equation}
 where $\dots$ refers to next to leading contributions in $\epsilon_2$. 
We point out that, in analogy with \eqref{simpleLIMIT},  only connected \emph{tree graphs} contributes to \eqref{equationTREE}.
As opposed to \eqref{simpleLIMIT},  now they are tree graphs on the set $[\ell]$ of $\ell$ clusters rather then the set of $k$ fundamental particles.
Notice that we did not expand the  $S_{\mathbf{Y}}$ factor. 
The integration in \eqref{Mayer} is still  over $k$ variables. 

The next step is to explicitly perform the integration over the distances of particles within the same cluster. This turns out to be essentially the same as in Section \ref{byregions}.
For each cluster $\mathbf{Y}_a$ we introduce a ``center of cluster'' coordinate $\overline{\phi}_a$ as in \eqref{introduceCenter}. 
It is clear that the $f_{IJ}$ entering the factors $S_{\mathbf{Y}_a}$, see \eqref{Sdef}, 
are independent of the  ``center of cluster'' coordinates $\overline{\phi}_a$.
Using  this observation and \eqref{equationTREE} we write 
\begin{align}
 \int\,\prod_{I=1}^k d\phi_I \,\,
\mathfrak{S}_{\{\mathbf{Y}_1,\dots,\mathbf{Y}_{\ell}\}}\qquad\qquad\qquad\qquad\qquad\qquad\qquad \nonumber \\
=\,\frac{1}{\epsilon_2}\,\int\,
\prod_{a=1}^{\ell}\left[d\overline{\phi}_a\,d\mu_{\mathbf{Y}_a}\prod_{I_a\,\in\,\mathbf{Y}_a}Q(\overline{\phi}_a+\phi_{I_a})\right]\,
\sum_{g\,\in\,\mathcal{T}^{[\ell]}_c}
\,\prod_{e(a,b)\,\in\, g}\,
\,\prod_{\substack{I_a\,\in\,\mathbf{Y}_a\\I_b\,\in\,\mathbf{Y}_b}}
G(\overline{\phi}_{ab}+\phi_{I_aI_b})+\dots
\label{resultMayer}
\end{align}
where
\begin{equation}
 d\mu_{\mathbf{Y}}\,:=\,\frac{1}{\epsilon^{|\mathbf{Y}|-1}_2}\,
\prod_{I\,\in \,\mathbf{Y}_a }\,\frac{d\phi_I}{2 \pi i }\,
\delta\left(\frac{1}{|\mathbf{Y}|}\sum_{I\,\in\,\mathbf{Y}}\,\phi_I\right)\,
\sum_{g\,\in\,\mathcal{G}^{\mathbf{Y}}_c}\,
\prod_{e(I,J)\,\in\, g}\,f_{IJ}\,.
\end{equation}
As extensively discussed in the previous sections, the leading contribution in $\epsilon_2$ of integrals of the type
\eqref{resultMayer}, can be obtained by neglecting the deviation from the center of cluster coordinate in the functions $Q(x)$ and $G(x)$. Doing so,   \eqref{resultMayer} becomes
\begin{equation}
 \int\,\prod_{I=1}^k d\phi_I \,\,
\mathfrak{S}_{\{\mathbf{Y}_1,\dots,\mathbf{Y}_{\ell}\}}\,=\,
\frac{1}{\epsilon_2}\,\int\,
\prod_{a=1}^{\ell}d\overline{\phi}_a\,Q^{|\mathbf{Y}_a|}(\overline{\phi}_a)\,\mathcal{J}_{|\mathbf{Y}_a|}\,
\sum_{g\,\in\,\mathcal{T}^{[\ell]}_c}
\,\prod_{e(a,b)\,\in\, g}\,|\mathbf{Y}_a|\,|\mathbf{Y}_b|\,
G(\overline{\phi}_{ab})+\dots
\label{resultMayer2}
\end{equation}
$\mathcal{J}_{k}$ was defined in \eqref{defJk} and computed in \eqref{resultJk}. For convenience we recall it here
\begin{equation}
 \mathcal{J}_{|\mathbf{Y}_a|}\,=\,\int\,d\mu_{\mathbf{Y}_a}
\,=\,\frac{1}{2\pi i}\,\frac{|\mathbf{Y}_a|!}{|\mathbf{Y}_a|^2}\,.
\end{equation}
In the limit of small $\epsilon_2$, the summands in \eqref{Mayer} depend only on the sizes $n_1,\dots,n_{\ell}\geq 1$ of the clusters corresponding to $\mathsf{b}\,\in\,\mathcal{B}^{[k]}$.
Converting the sum in \eqref{Mayer} from  a sum over groupings of $k$ particles to a sum over the number of clusters  $\ell$ 
and their sizes $n_1,\dots,n_{\ell}$ produces a factor $\frac{k!}{\ell! n_1!\,\dots\,n_{\ell}!}$.
Assembling the pieces together we finally arrive at the main result 
\begin{equation}
 \lim_{\epsilon_2\to 0}\epsilon_2\,\log\,\mathcal{Z}\,=\,\sum_{\ell=1}^{\infty}\frac{1}{\ell!}\,\sum_{n_1,\dots,n_{\ell}\geq 1}\,
\int\,\prod_{a=1}^{\ell}\frac{d\phi_a}{2\pi i}\,\frac{q^{n_a}\,Q^{n_a}(\phi_a)}{n_a^2}\,
\sum_{g\,\in\,\mathcal{T}_c^{[\ell]}}\,\prod_{e(a,b)\,\in\,g}\,n_a\,n_b\,G(\phi_{ab})\,,
\label{mainresultfromMayer}
\end{equation}
where we replaced $\overline{\phi}_a$ with $\phi_a$.
We emphasize once again  that the sum has been rearranged from a sum over particles, weighted by $q$, to a sum over clusters.
 More precisely, for each  $\ell$ there are $\ell$ clusters and $\ell-1$ $G$-factors.
For $\ell=1$ one immediatly recovers the dilogarithm
\begin{equation}
\int\,\frac{d\phi}{2 \pi i }\,\text{Li}_2\left(q\,Q(\phi)\right)) \,.
\label{Li2resum}
\end{equation}
 For $\ell=2$ one has 
\begin{equation}
\frac{1}{2}\,\int\,\frac{d\phi_1}{2 \pi i }\frac{d\phi_2}{2 \pi i }\mathcal{Q}_{\text{eff}}(\phi_1)\,\mathcal{Q}_{\text{eff}}(\phi_2)\,G(\phi_1-\phi_2) \,,
\qquad
\mathcal{Q}_{\text{eff}}(\phi)\,=\,\log(1-q\,Q(\phi))\,.
\label{effectiveresum}
\end{equation}

The result \eqref{mainresultfromMayer} can be visualized  diagrammatically in a simple way.
For each $\ell$ in the sum one draws all three graphs on the set $\{1,2,\dots,\ell\}$.
To each node $a$ of the graph is associated an integer $n_a\geq 1$.
Once the diagrams are drawn, the corresponding integrals can be written using the following
``Feynann rules''
\begin{itemize}
 \item Vertex
\begin{equation}
    \begin{tabular}{c}{\includegraphics[height=.9cm]{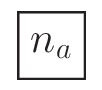}}\end{tabular}  \,= \ \,\frac{q^{n_a}}{n_a^2}\int\frac{d\phi_a}{2\pi i }\, Q^{n_a}(\phi_a)\,
\end{equation}
\item Propagator
\begin{equation}
    \begin{tabular}{c}{\includegraphics[height=.9cm]{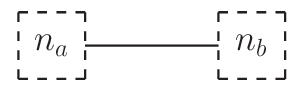}}\end{tabular}  \,= \ \, n_a\,n_b\,G(\phi_{ab})\,.
\end{equation}
Here we used a dashed square to indicate that the propagator is associated to the edge only.
\end{itemize}
The graphs contributing to  $\ell=1,2,3$ are given by
\bea
\ell=1\,& : & \ \ \ \ \, \sum_{n_1=1}^{\infty}\,   \begin{tabular}{c}{\includegraphics[height=.8cm]{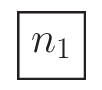}}\end{tabular}
\\
\ell=2\,& : & \ \ \, \sum_{n_1,n_2=1}^{\infty}\,   \begin{tabular}{c}{\includegraphics[height=.8cm]{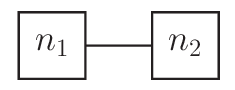}}\end{tabular}
\\
\ell=3\,& : & \ \sum_{n_1,n_2,n_3=1}^{\infty}\,  \begin{tabular}{c}{\includegraphics[height=.8cm]{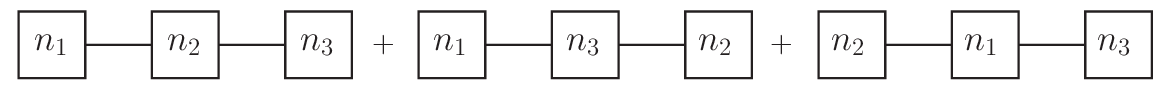}}\end{tabular}
\eea
The first two graphs correspond to \eqref{Li2resum} and \eqref{effectiveresum} respectively.
Notice that each integral obtained by applying the Feynman rules, should be multiplied by  an overall  $\frac{1}{\ell!}$ factor, which is explicit in \eqref{mainresultfromMayer}.
Graphs with more vertices, upon dressing the vertices with positive integers, are as in Figure \ref{connect_tree_graph}.

\section{Derivation of the  TBA}

In this section we show that the logarithm of the instanton partition function in the $\epsilon_2\rightarrow 0 $ limit, whose structure have been studied in last section,
 satisfies TBA equations.
 We will first provide a perturbative check to a very high order in the instanton counting parameter $q$, and then present an all order proof based on rewriting the instanton partition function as a $(0+1)$-dimensional path integral.

\subsection{Perturbative expansion \label{perturbative_TBA}}

Let us recall the expression of the superpotential coming from the TBA
\be \calW =  \int {dx \over 2\pi i} \Big[ -{1\over2} \varphi(x) \log\left(1 - q Q(x) e^{-\varphi(x)} \right)  + {\rm Li}_2\left( q Q(x) e^{-\varphi(x)}\right) \Big] \,,  \ee
where
\be \varphi(x) = \int {d y \over 2\pi i} \, G(x-y) \log\left( 1-qQ(y) e^{-\varphi(y)} \right)\,, \ee
compare to  \eqref{tbaW} and \eqref{tbaphi}.
The basic observation is that TBA equation has a natural expansion in terms of tree graphs, therefore one can compare it directly with the Mayer expansion of the instanton partition function \eqref{mainresultfromMayer}. 

One writes $\varphi$ and ${\cal W}$ as
\be \varphi(x) = \sum_{k=1}^\infty q^k \,\varphi_k(x) \,, \qquad \calW = \sum_{k=1}^\infty q^k \,\calW_k \, . \ee
$\varphi$ can be solved recursively via the TBA equation, for example up to the first two orders
\bea \varphi_1(x) &=& - \int {d y \over 2\pi i} G(x-y)  Q(y) \, , \\ 
\varphi_2(x) &=&  - {1\over2} \int {d y \over 2\pi i} G(x-y)  Q(y)^2 -  \int {d y \over 2\pi i} {d z \over 2\pi i} G(x-y) G(y-z) Q(y) Q(z)  \,.  \eea
Substituting the expression for  $\varphi$ in ${\cal W}$ and collecting the terms at a given order of $q$, one obtains, here up to order $q^3$
\bea \calW_1 &=& \int {d x \over 2\pi i} \, Q(x) \, , \\ 
\calW_2 &=& {1\over 4}\int {d x \over 2\pi i} \, {Q(x)^2} + {1\over2} \int {dx \over 2\pi i} {dy \over 2\pi i} \, {Q(x) G(x-y) Q(y)} ,  \\ 
\calW_3 &=& {1\over 9} \int {d x \over 2\pi i} {Q(x)^3} + {1\over2} \int {dx \over 2\pi i} {dy \over 2\pi i} \, {Q(x)^2 G(x-y) Q(y)}  \nonumber\\ && + {1\over2} \int {dx \, dy \, dz \over (2\pi i)^3} \, {Q(x) G(x-y) Q(y) G(y-z) Q(z)} \,. \eea
One can see that the expansion of $\calW$ has a natural interpretation in terms of connected tree diagrams, with $G(x-y)$ as propagators and $Q(x)^i$ as vertices.
This is the same structure as the one obtained starting from the integral representation of the instanton partition function, see \eqref{mainresultfromMayer}.
The  non-trivial thing to show is that not only the structure is similar but that the  coefficients actally match.

By staring at their form, it is not obvious that the Mayer expansion \eqref{mainresultfromMayer} and the TBA actually produce the same result.
In particular to obtain the contribution of a single graph in  \eqref{mainresultfromMayer} one needs to take into account a large number of terms coming from the  $\varphi$ expansion.
This procedure can be  straightforward implemented on a compute. We have checked up to $q^7$ order and found perfect agreement with the Mayer-Cluster expansion of the instanton partition function.
 It is  remarkable that these two different structures are actually equivalent with each other.
 In the next subsection, we will see the equivalence of the two expansions can be  proven to all order in $q$.

\subsection{TBA as saddle-point of a $(0+1)$-dimensional path integral \label{sec:nonpert}}

As we will shortly review, there is a nice way to rewrite a grand canonical partition function of the form \eqref{Zinst} as a path integral, based on a trick of Polyakov \cite{Polyakov}.
Path integrals are particularly well suited to be studied in the classical limit $\hbar\rightarrow 0$, identified with $\epsilon_2$ in this section, where the path integral is obtained by evaluating
 the classical action on some solution to the equations of motion.
In the case of  the instanton partition function \eqref{Zinst}, this limit is more subtle as the two particle potential  in the original statistical mechanical problem is singular in this limit.
Using the considerations from section \ref{originDILOG} we will be able to disentangle such contributions and derive the TBA in a rather transparent way as a saddle point equation of an appropriate action.

Let us start by reviewing the steps presented in the appendix of  \cite{Polyakov}. Consider the partition function
\be Z = \sum_{k=0}^\infty {q^k \over k!}  \int   \prod_{I=1}^k {d \phi_I \over 2\pi i} \prod_{I < J}^k  \, e^{-V(\phi_{IJ})}  \,. \label{ZforPolyakov}\ee
%
The basic idea is to rewrite the interactions as 
\be   \prod_{I<J}^k e^{-V(\phi_{IJ})}\,=\,e^{\Delta({\delta\over\delta \varphi})} \,
 \prod_{I=1}^k e^{-\varphi(\phi_I)} \Big|_{\varphi\rightarrow0}
 \,,
\label{interPolya} \ee
where
\be \Delta\Big({\delta\over\delta \varphi}\Big) = - {1\over2} \int d x \, dy\, V(x-y) {\delta \over \delta \varphi(x)} {\delta \over \delta \varphi(y)} 
\,.\ee
The operator $e^{\Delta}$ can be pulled out from the sum over the number of particles $k$ to rewrite
\be Z = e^{\Delta({\delta\over\delta \varphi})} \bigg( \sum_{k=0}^\infty {1 \over k!} \int  \prod_{I=1}^k {d \phi_I \over 2\pi i}  e^{-\varphi(\phi_I)}  \bigg) \bigg|_{\varphi=0} = e^{\Delta({\delta\over\delta \varphi})} \exp\left( \int {d\phi_0 \over 2 \pi i} \, { e^{-\varphi(\phi_0)}}\right) \bigg|_{\varphi\rightarrow 0}\ . \ee
%
All the information about the interaction in the original gas \eqref{ZforPolyakov} is now encoded in the operator $\Delta$.
To translate the construction above into path integral language one observes the following equality
\be
e^{\Delta({\delta\over\delta \varphi})}\,\mathsf{F}[\varphi]\Big{|}_{\varphi=0}\,=\,
\int [ D \varphi\, D\rho  ]\, e^{\frac{1}{\hbar} \left[ \hbar\,\Delta(\frac{i \rho}{2\pi \hbar})+\int\frac{dx}{2\pi i }\rho(x)\,\varphi(x)\right]}\,\mathsf{F}[\varphi]\,.
\label{fromDeltatoPATH}
\ee
This can be formally proved by Fourier transformation\,%
\footnote{The relation \eqref{fromDeltatoPATH} is a path integral analogue of  
\be e^{\Delta({\partial \over \partial x})} \, \mathbf{f}(x)\big|_{x\rightarrow0} =  \int {d p \over 2\pi } \, e^{ i p x + \Delta(i p)} {\hat {\mathbf{f}}}(p)\big|_{x\rightarrow0} =  \int {d p dy \over 2\pi} \, e^{-i p y + \Delta(i p)} {\mathbf{f}}(y)  \, , \label{FTtrick} \ee
where ${\hat {\mathbf{f}}}(p)$ is introduced via Fourier transformation.}. 
The measure in $[ D \varphi\, D\rho  ]$ is defined  in such a way that if $\mathsf{F}[\varphi]=1$ then the resulting path integral should also evaluate to $1$.
Notice that the relation \eqref{fromDeltatoPATH} holds also when the operator $\Delta$ is not quadratic.
 If $\Delta$ is linear $\Delta=\int_x \gamma(x){\delta\over\delta \varphi(x)}$, the $\rho$ integration produces a delta function and the resulting path integral gives $\mathsf{F}[\gamma]$ as it should.
Indeed in this case $e^{\Delta}$ is just the shift operator.

From the discussion in Section \ref{sec:Mayer-Cluster}, it is rather  clear that, in the study of the instanton partition function, we should apply the procedure just described in a sightly modified way.
In particular we will factor out only the long range part of the potential \eqref{splittingPOT}, corresponding to $\widetilde{\mathcal{D}}$, as we will now describe.
The first step is to introduce 
\bea \Delta_{\rm L} \Big({\delta\over\delta \varphi}\Big)= - {1\over2} \int d x \, dy\,  V_{\rm L}(x-y) {\delta \over \delta \varphi(x)} {\delta \over \delta \varphi(y)}\,,
\qquad e^{-V_{\rm L}(x)}\,=\,\widetilde{\mathcal{D}}(x)  \,.
\eea
As above, one uses this operator to rewrite the long range interactions as in \eqref{interPolya}, and  factor out their contribution from the sum over the instanton number $k$  as follows 
\be {\cal Z} = e^{\widetilde\Delta({\delta\over\delta \varphi})}\, \mathbf{F}[\varphi] \Big|_{\varphi=0} \,,
\qquad 
 \mathbf{F}[\varphi] = \sum_{k=0}^\infty {1 \over k!} \int  \prod_{I=1}^k {d \phi_I \over 2\pi i}{q Q(\phi_I) e^{-\varphi(\phi_I)} \over \epsilon_2} \prod_{I < J}^k {\phi_{IJ}^2 \over \phi_{IJ}^2 - \epsilon_2^2}\,.\label{Zgood}
\ee
Notice that we did not expand the short range interaction ${\phi_{IJ}^2 \over \phi_{IJ}^2 - \epsilon_2^2}$, which still appears in $\mathbf{F}[\varphi]$.
The reader may recognize that we already calculated this integral in the  $\epsilon_2\rightarrow 0$ limit in Section \ref{originDILOG}. 
More precisely  $\mathbf{F}[\varphi]$ coincides with $\mathcal{Z}_{\text{Short}}$ defined in   \eqref{ZwithF=0}, upon replacing  $Q(x)$ with  $Q(x) e^{-\varphi(x)}$.
Using the result \eqref{resultLi2} we can thus rewrite
\be
\mathbf{F}[\varphi] \,=\,\exp\left(\frac{1}{\epsilon_2}\, \int {d\phi_0 \over 2 \pi i} \,  {\rm Li}_2[qQ(\phi_0)\, e^{-\varphi(\phi_0)}]+O(\epsilon_2^0)\right)\,.
\label{Li2good}
\ee
 The appearance of the dilogarithm is quite remarkable. 
The derivation presented makes its origin from short range interactions completely transparent. 
The expression \eqref{Zgood} with \eqref{Li2good}, is well suited for the semiclassical analysis in the $\epsilon_2\rightarrow 0 $ limit. 
Using \eqref{fromDeltatoPATH} in this case, and recalling that  $V_{\rm L}(x)=- \epsilon_2 G(x) + {\cal O}(\epsilon_2^2)$ one immediatly obtains
\be
\mathcal{Z}\,=\,\int [ D \varphi\, D\rho ]\, \exp \left[ { \frac{1}{\epsilon_2}\left(\mathcal{Y}[\rho,\varphi]+O(\epsilon_2)\right)} \right] \,, 
\label{Zfinal}\ee
where $\mathcal{Y}[\rho,\varphi]$ is defined in \eqref{YYdef} and
 we identified $\hbar$ with $\epsilon_2$.  In the limit of small $\epsilon_2$ this path integral is calculated by the sadle point method\,%
 \footnote{Alternatively, one can also integrate out $\rho$ exactly, as it appears quadratically  in the action, before taking the $\epsilon_2\rightarrow 0 $ limit.
 The resulting path integral, in $\varphi$ in the limit of small $\epsilon_2$,  is dominant by the saddle point of $\varphi$, where the saddle point equation is exactly the TBA equation \eqref{tbaphi}.}. 
The superpotential is thus obtained from the critical valeus of the functional  $\mathcal{Y}[\rho,\varphi]$. As explained in Section  \ref{sec:NS},  the TBA immidiately follows. This complets the derivation.

\subsubsection{Multi-particle interactions}

Let us mention that
the above discussion may be generalized to the cases containing ``multi-particle" interactions, such as
\be \mathcal{Z} = \sum_{k=0}^\infty {q^k \over k!}  \int   \prod_{I=1}^k {d \phi_I \over 2\pi i}{q Q(\phi_I) \over \epsilon_2} \prod_{I < J}^k {\phi_{IJ}^2 \over \phi_{IJ}^2 - \epsilon_2^2} \prod_{m=2}^\infty \prod_{I_1 < \ldots< I_m}^k  \, e^{-V_m(\phi_{I_1}, \phi_{I_2}, \ldots, \phi_{I_m})} \,. \ee
One can similarly introduce
\bea {\Delta_m} \Big( {\delta\over\delta \varphi} \Big) &=& - {1\over m!} \int d^m x \, V_m(x_1,...,x_m)  \prod_{I=1}^m {\delta \over \delta \varphi(x_I)}  \, . \eea
In analogy with \eqref{fromDeltatoPATH}, this can be brought to the following form
\be \mathcal{Z} =  \int [D \rho D \varphi] \, \exp\left[ - \sum_{m=2}^\infty {(-1)^m\over m!} \int {d^m x \over (2\pi i)^m} \, V_m(x_1,..,x_m)  \prod_{I=1}^m {\rho(x_I) \over \epsilon_2} + { \int {d x \over 2\pi i \,\epsilon_2}  \rho(x) \varphi(x)}   \right]  {\rm {\bf F}}[ \varphi] . \label{multi2} \ee
Notice that if $m>2$ one can no longer integrate out $\rho$ exactly. 
The tri-particle interaction is directly related to certain gauge theories. The main example is given by  quiver theories containing matter in the trifundamental representation of $U(2)^3$ \cite{Gaiotto:2009we}. 
The contour integral representation of the instanton partition function for this case can be found in   \cite{Hollands:2011zc}. 
Notice that one has $V_2=-\epsilon_2\,G+{\cal O}(\epsilon_2^2)$, $V_3=\epsilon_2^2\,G_3+{\cal O}(\epsilon_2^3)$. This is the right scaling to produce a $\frac{1}{\epsilon_2}$ factor in the exponent of \eqref{multi2} which is crucial for the semiclassical analysis.


\section{Generalization to quiver gauge theories \label{sec:quiver}}

Until now the object of  study has been the instanton partition function for pure $\mathcal{N}=2$ 
Super-Yang-Mills with gauge group $SU(N)$, see \eqref{Zinst}.  
This partition function admits a natural generalization corresponding to other 
$\mathcal{N}=2$ gauge theories. 
In order to specify an $\mathcal{N}=2$ supersymmetric gauge theory the first step is to fix the following data: a gauge group $G$ and a set of irreducible representations $R$ of $G$  corresponding to its matter content.
Such  datum gives rise to a  consistent, or UV-complete, theory only if the beta function $\beta_\mathsf{v}$ associated to each simple factor of $G=\prod_{\mathsf{v}}\,G_\mathsf{v}$ satisfies $\beta_\mathsf{v} \leq 0$.  
 This requirement puts strong constraints on $R$ and $G$. A classification of consistent theories can be found in the recent work \cite{Bhardwaj:2013qia}. In particular, with the exception of $G=SU(2)^V$ with $R$ given by trifundamentals introduced in \cite{Gaiotto:2009we}, 
each irreducible component of $R$ is charged under at most two $G_\mathsf{v}$. In the following we restrict to the case 
\begin{equation}
 G\,=\,\prod_{\mathsf{v}=1}^V\,SU(N_\mathsf{v})\,,
\qquad
R\,=\,\sum_{\mathsf{v},\mathsf{w}} c_{\mathsf{v}\,\mathsf{w}}\, (\square_\mathsf{v}, \overline{\square}_\mathsf{w}) + \sum_\mathsf{v} \,n_\mathsf{v} \square_\mathsf{v}\,.
\label{gaugematter}
\end{equation}
Theories in this family with vanishing beta functions
falls into three classes, referred to as type  I, II, II* in \cite{Nekrasov:2012xe}.
The matrix $c$ in \eqref{gaugematter} is then identified
with the adjacency matrix of the Dynkin diagram of type $ADE$ (type I)
and $\widehat{A}\widehat{D}\widehat{E}$ (type II and II*). 
Any other consistent theory in the family \eqref{gaugematter} can be obtained  as appropriate limit of  theories of type I, II, II*.

We fix $G$ and $R$  as in  \eqref{gaugematter}.  We denote respectively by $\mathsf{V}$ and $\mathsf{E}$ the set of vertices and edges of the graph represented by $c_{\mathsf{v}\,\mathsf{w}}$. We denote the number of vertices by $V=|\mathsf{V}|$.
 The instanton partition function depends on the following quantities
\begin{itemize}
 \item $\Omega$-deformation parameters $\epsilon_1$ and $\epsilon_2$, we set $\epsilon=\epsilon_1+\epsilon_2$
\item gauge couplings $q_{\mathsf{v}}$,   $\mathsf{v} \in \mathsf{V}$
\item Coulomb parameters $a\in$ Cartan subalgebra of $\text{Lie}(G)$, $a_{\mathsf{v},\lambda}$ for  $\mathsf{v}  \in \mathsf{V}$, $\lambda=1,\dots,N_{\mathsf{v}}$
\item fudametal masses $m_{\mathsf{v},f}$, $\mathsf{v}\in \mathsf{V}$, $f=1,\dots,n_\mathsf{v}$
\item bifundamental masses\footnote{To avoid confusion we recall that in our conventions a bifundamental hypermultiplet
associated to the edge $\mathsf{e}$ corresponds to a factor
$(\square_{s_{\mathsf{e}}},\,
 \overline{\square}_{t_{\mathsf{e}}})+(\square_{t_{\mathsf{e}}}, \,\overline{\square}_{s_{\mathsf{e}}})$ 
in \eqref{gaugematter}.} $m_\mathsf{e}$, $\mathsf{e} \in \mathsf{E}$.
\end{itemize}
We package masses and Coulomb parameters into polynomials as follows
\be M_\mathsf{v}(x) = \prod_{f=1}^{n_\mathsf{v}} (x - m_{\mathsf{v},f}) \,,
 \qquad P_\mathsf{v}(x) = \prod_{\lambda=1}^{N_\mathsf{v}} (x - a_{\mathsf{v},\lambda}) \, . \ee
The contour integral representation of the Nekrasov partition function takes the form
\be {\cal Z} = 
\sum_{k_1,\, \ldots,\,k_V}^\infty\,
\left(\frac{\epsilon}{\epsilon_1\,\epsilon_2}\right)^{k}\,
\frac{q^{k_1}_1}{k_1!}\,\dots\,\frac{q^{k_V}_V}{k_V!}\,
\int\,\prod_{\mathsf{v}\,\in \,\mathsf{V}}\,
\Big{(}[d\phi_\mathsf{\mathsf{v}}] \, z_{\mathsf{v},\,k} ^{\rm gauge}(\phi_\mathsf{v})\, z_{\mathsf{v},\,k}^{\rm fund}(\phi_\mathsf{v})\Big{)}\,
\prod_{\mathsf{e}\,\in\,\mathsf{E} }\,
 z_{\mathsf{e},\,k}^{\rm bifund}(\phi_{s_{\mathsf{e}}},\phi_{t_{\mathsf{e}}})\,
\label{Zquiver} 
\ee
where $k=\sum k_\mathsf{v}$ and $[d\phi_\mathsf{v}]=\prod_{I=1}^{k_\mathsf{v}}(2\pi i)^{-1} d\phi_{\mathsf{v},I}$.
As above, $\mathsf{V}$  and $\mathsf{E}$ are respectively  set of vertices and edges of the quiver. Moreover, given an edge $\mathsf{e}$, its orientation defines two vertices called ${s_{\mathsf{e}}}$  (source) and $t_{\mathsf{e}}$ (target).
Using the expressions for $z_\mathsf{v} ^{\rm gauge}$, $z_\mathsf{v}^{\rm fund}$ and  $z_{\mathsf{e}}^{\rm bifund}$
given in Appendix \ref{app:quiverblocks} we rewrite the integrand of \eqref{Zquiver} as 
\begin{equation}
 \prod_{\mathsf{v}\,\in \,\mathsf{V}}
\,\left[
 \left(\prod_{I=1}^{k_{\mathsf{v}}} \,Q_{\mathsf{v}}(\phi_{\mathsf{v},I})\right)\,
 \left(  \prod_{1\leq I < J\leq\, k_\mathsf{v}}{\cal D}(\phi_{\mathsf{v},I}-\phi_{\mathsf{v},J})  \right)\right]\,
 \prod_{\mathsf{e}\,\in \,\mathsf{E}}
\,\left[
 \prod_{I=1}^{k_{\mathsf{s_{\mathsf{e}}}}}\prod_{J=1}^{k_{t_{\mathsf{e}}}}\,\mathcal{D}_{\mathsf{e}}(\phi_{s_{\mathsf{e}},I}-\phi_{t_{\mathsf{e}},J}) \right]\,
\label{ZquiverINT} 
\end{equation}
where
\begin{equation}
 Q_{\mathsf{v}}(x)  \,=\, 
{ M_{\mathsf{v}}(x) 
\over P_{\mathsf{v}}(x)\, P_{\mathsf{v}}(x+ \epsilon)}\,
\prod_{\mathsf{v}\stackrel{\mathsf{e}}{\rightarrow}\mathsf{v}'} P_{\mathsf{v}'}(x - m_{\mathsf{e}})\,
\prod_{\mathsf{v}\stackrel{\mathsf{e}}{\leftarrow}\mathsf{v}'} P_{\mathsf{v}'}(x + m_{\mathsf{e}}+\epsilon)  \, ,
\label{Qdef}
\end{equation}
\begin{equation}
 \mathcal{D}(x)\,=\,\Delta(x)\,\Delta(-x)\,,\qquad
 \mathcal{D}_{\mathsf{e}}(x)\,=\, [\Delta(m_{\mathsf{e}}+x)\,\Delta(m_{\mathsf{e}}-x)]^{-1}\,,
 \qquad
\Delta(x)\,=\,\frac{x\,(x+\epsilon_1+\epsilon_2)}{(x+\epsilon_1)\,(x+\epsilon_2)}\,,
\label{Ddef}
\end{equation}
Note that $\mathcal{D}$ is the same as for pure SYM, see  \eqref{defD}.
Moreover, we have the manifest symmetry
${\cal D}(x) = {\cal D}(-x)$ and ${\cal D}_{\mathsf{e}}(x) = {\cal D}_{\mathsf{e}}(-x)$.

Explanations regarding the origin of this expression are collected in Appendix \ref{app:contour_integral}. The expression for the contribution to the partition function from bifundamental matter can be found in  \cite{Shadchin:2005cc} and \cite{Billo:2012st}. For more general gauge groups and representation see e.g.~\cite{Hollands:2010xa,Hollands:2011zc} and references therein.

\subsection{TBA for quiver gauge theories}

The partition function \eqref{Zquiver} in the limit $\epsilon_2\rightarrow 0$ can be studied in a similar way as for the simpler case of pure $SU(N)$ SYM, see \eqref{Zinst}. In particular it can be characterized as solution of  certain TBA equations given below. 
The important observation to be made is that as long as the bifundametal masses $m_{\mathsf{e}}$ are large compared to $\epsilon_2$,  the pair-wise interaction term in the non-ideal gas interpretation of \eqref{ZquiverINT} 
splits into two terms with different scales, compare to \eqref{splittingPOT}. 
The first term  ${x^2 \over x^2 -\epsilon^2_2}$,  present for each gauge group, gives rise to the dilogarithm function in the TBA. The remaining factor in the pair-wise interaction is responsable for the kernel term in the TBA non-liear integral equation. 

The resulting TBA is summarized as follows.
As in \cite{Nekrasov:2009rc}, the partition function can be written as the critical value of the Yang-Yang functional
\be {\cal Y}(\rho, \varphi) =  {1\over2} \sum_{\mathsf{v},\mathsf{w}} \int \frac{d x}{2 \pi i }\,\frac{d y}{2 \pi i } \, \rho_\mathsf{v}(x) G_{\mathsf{v} \mathsf{w}}(x-y) \rho_\mathsf{w}(y)  + \sum_\mathsf{v} 
\int  \frac{d x}{2 \pi i }\Big[ \rho_\mathsf{v}(x) \varphi_\mathsf{v}(x) + {\rm Li}_2\left( q_\mathsf{v} Q_r(x) e^{-\varphi_\mathsf{v}(x)}\right) \Big]  , \ee
where the variations with respect to $\varphi_\mathsf{v}(x)$ and $\rho_\mathsf{v}(x)$ give
\be  \rho_\mathsf{v}(x) = - \log\left[ 1-q_\mathsf{v} \, Q_\mathsf{v}(x) e^{-\varphi_\mathsf{v}(x)} \right] , 
\qquad \varphi_\mathsf{v}(x) = - \sum_\mathsf{w} \int \frac{d y}{2 \pi i } \, G_{\mathsf{v} \mathsf{w}}(x-y) \rho_\mathsf{w}(y) \, . \ee
The twisted superpotential for the quiver gauge theory is obtained as
\be {\cal W} = \sum_\mathsf{v} \int \frac{d x}{2 \pi i } \Big[ -{1\over2} \varphi_\mathsf{v}(x) \log\left(1 - q_\mathsf{v}\, Q_\mathsf{v}(x) e^{-\varphi_\mathsf{v}(x)} \right)  + {\rm Li}_2\left( q_\mathsf{v} \, Q_\mathsf{v}(x) e^{-\varphi_\mathsf{v}(x)}\right) \Big] , \ee
where $\varphi_\mathsf{v}(x)$ satisfy the TBA equation
\be \varphi_\mathsf{v}(x) = \sum_\mathsf{w} \int \frac{d y}{2 \pi i } \, G_{\mathsf{v} \mathsf{w}}(x-y) \log\left[ 1-q_\mathsf{w} \, Q_\mathsf{w}(y) e^{-\varphi_\mathsf{w}(y)} \right] .
 \ee
The various functions which contain the data of quiver theory are
\be  G_{\mathsf{v}\mathsf{w}}(x) = {d \over d x} \left[ 
 \delta_{\mathsf{v}\mathsf{w}} \log \left( { x+ \epsilon_1 \over x- \epsilon_1} \right) +
 c_{\mathsf{v}\mathsf{w}}  \log \left( { x+ m_{\mathsf{v}\mathsf{w}} \over x + m_{\mathsf{v}\mathsf{w}} + \epsilon_1} \right) 
+ c_{\mathsf{w}\mathsf{v}}  \log \left( { x - m_{\mathsf{w}\mathsf{v}}-\epsilon_1  \over x - m_{\mathsf{w}\mathsf{v}}} \right) \right] \, , 
\label{Rvw}
\ee
and  $Q_{\mathsf{v}}(x)$ defined  in \eqref{Qdef}.
In \eqref{Rvw} we defined $m_{\mathsf{v}\mathsf{w}}:=m_{\mathsf{e}}$ if the vertices $\mathsf{v}$ and $\mathsf{w}$ are connected by the edge $\mathsf{e}$ and $m_{\mathsf{v}\mathsf{w}}:=0$ otherwise.   
The expression for the propagator $G_{\mathsf{v}\mathsf{w}}(x)$ given above is obtained as as
\be
 G_{\mathsf{v}\mathsf{w} }(x) = 
\begin{cases}
 {\rm Limit}_{\epsilon_2 \rightarrow0}
 {{\cal D}(x) - 1 \over \epsilon_2} & \text{for}\,\,\,\mathsf{v}=\mathsf{w} \\
{\rm Limit}_{\epsilon_2 \rightarrow0} {{\cal D}_{\mathsf{e}}(x) - 1 \over \epsilon_2} & \text{for}\,\,\,\mathsf{v}\stackrel{\mathsf{e}}{\text{\----}}\mathsf{w}  \\
0 & \text{otherwise}
\end{cases}
\ee
Note that  in this definition we have $G_{\mathsf{v}\mathsf{w}}(x) = G_{\mathsf{w}\mathsf{v}}(-x)$.


\section{Conclusion and Discussion \label{sec-discussion}}

In this paper we studied the instanton partition functions of four dimensional ${\cal N}=2$ gauge theories in a special limit of $\Omega$ deformation parameters, namely, taking $\epsilon_2 \rightarrow0$ but keeping $\epsilon_1$ finite.  We show explicitly that the instanton part of the  twisted superpotential $\mathcal{W}$, which in statistical mechanics language is equal to the free energy, satisfies TBA equations.
It is thus naturally identified with the Yang-Yang function of some quantum integrable model \cite{Nekrasov:2009rc}.
Based on the proof, we are also able to generalize the correspondence to a general class of ${\cal N}=2$ quiver theories. In this case one obtains a set of coupled TBA equations with one equation for each node of the quiver.

The starting point of the derivation is the contour integral representation of the Nekrasov partition function. It can be interpreted as the grand canonical partition function of a non-ideal gas of particles.
Due to the singular structure  of the two-particles interaction, an interesting  effective description of this gas emerges  in the $\epsilon_2\rightarrow 0$ limit.  
In this effective description the ``new particles" are clusters of the original particles and are subject to certain tree-level interaction only.
This picture is best obtained by employing the so-called iterated Mayer expansion \cite{Gopfert:1981zu}.
This structure is nicely captured by the TBA, in particular the clustering of particles and their effective interaction is nicely reproduced by the presence of
the Li$_2$ function\footnote{This makes it drastically different from some other TBA obtained by simply applying saddle point method \cite{Poghossian:2010pn, Fucito:2011pn, Fucito:2012xc}. See also \cite{Bourgine:2013ipa}.}.
An all order proof in the instanton counting parameter $q$ of the TBA equations is given applying a $(0+1)$-dimensional path integral trick, dating back to Polyakov \cite{Polyakov}.

In studying the  NS limit of the instanton partition function, one has to face the problem of evaluating integrals when some parameter, in our case $\epsilon_2$, is small.
There is a systematic way to do so, particularly successful for calculating Feynman integrals, called the  method of expansion by regions.
While our results are obtained also independently of this method, it provides a simple way to single out the leading $\frac{1}{\epsilon_2}$ behavior of the logarithm of the partition function.
This structure is somewhat reminiscent of the exponentiation of infrared divergences in gauge theories. It would be interesting to further study the applicability of this method to the type
of integral considered here.

In this paper we considered  quiver gauge theories involving only $U(N)$ gauge group factors and matter in the fundamental/antifundamental or bifundamental representations.
It is natural to extend this work by deriving TBA equation in the case of the  other classical gauge groups $SO(N)$, $SP(2N)$ and to other matter content, e.g.~symmetric and antisymmetric representations of  $U(N)$ or trifundamental of $U(2)^3$.
Other interesting  generalizations are the deformation of the partition function by chiral ring operators and the five dimensional lift of the partition funtion.  The details of these studies will be reported elsewhere.

One of the most challenging question for the future is to understand how the proposed TBA equations emerge from the quantum integrable model point of view. 
In the case of  pure $SU(N)$  ${\cal N}=2$ super Yang-Mills,  which corresponds to the  periodic Toda chain with $N$ sites, this has been achieved in \cite{Kozlowski:2010tv}.
In the derivation a crucial role is played by the so-called  Baxter Q-operator, which is currently the  most powerful and universal tool for determining the spectrum of quantum integrable systems.
Despite successful applications in many cases, see \cite{Frenkel:1308} (and also \cite{Nekrasov:2013xda}) for  recent progress and relevant references, a complete theory of  Q-operators is still to be established. 
In particular a systematic construction is currently not available
in the case for which the representation of the relevant quantum group is not of highest-weight type, see \cite{MT} and references therein for some progress in this direction. 
Via the NS proposal, quantum integrable systems of this type can be systematically solved by gauge theory methods, offering an entirely new perspective on the structure underlying integrable models.

TBA equations of a similar type also appeared in the context of wall crossing effects for the BPS spectrum in ${\cal N}=2$ gauge theories \cite{Gaiotto:2008cd, Alexandrov:2010pp}. It would be interesting to study the connection with our work and find the quantum integrable models corresponding to the TBA appearing there. Another connection is to the study of minimal surfaces in AdS \cite{Alday:2009dv,Alday:2010ku}. 
In this case, the area of such surfaces is equal, via the AdS/CFT correspondence, to the strong coupling limit of the logarithm of null Wilson-loop expectation values in ${\cal N}=4$ SYM.
In \cite{Basso:2013vsa,Basso:2013aha}, it  has been proposed that the exact expectation value can be written as a sum of multiple integrals involving some basic building blocks called pentagon transition functions.
This form is structurally identical to the contour integral representation of the Nekrasov partition function. Moreover, the strong coupling limit correspond to the NS limit with $\frac{1}{\sqrt{\lambda}}\sim \epsilon_2$.  
Techniques of the type presented in this paper can be also used to show, directly from this representation of the Wilson loop,  that in the strong coupling limit it satisfies the TBA equations derived from the analysis of classical strings in AdS \cite{BassoTalk,BSV}.
It would be very interesting to study the possible connection with our work in this respect.


\section*{Acknowledgements}

We are greatly indebted to J\"{o}rg Teschner for suggesting this problem and valuable discussions at various stages of the project. We would like to thank Yasuyuki Hatsuda and Elli Pomoni for very useful discussions and collaboration on related topics. We would also like to thank Marcos Mari$\tilde{\rm n}$o for his interest in this work and encouragement.  
An important part of the work was done when G.Yang was a postdoc in the University of Hamburg where he was supported by the German Science Foundation (DFG) within the Collaborative Research Center 676 ``Particles, Strings and the Early Universe". C. Meneghelli is partially supported by a DFG grant in the framework of the SFB 676 ``Particles, Strings, and the Early Universe". G.Yang is supported by a DFG grant in the framework of the SFB 647 ``Space-Time-Matter".


\appendix

\section{On the  integral representation form of the Nekrasov partition function \label{app:contour_integral}}

In this Appendix we present a brief, non-self-consistent review, 
together with a collection of references, on the origin of the contour integral form of the instanton partition function \eqref{Zinst}, \cite{Moore:1997dj,Moore:1998et}.

The first step is to localize the path integral for the given $\mathcal{N}=2$ supersymmetric gauge theory  
in the $\Omega$-background to configuration of fields satisfying  the self-duality equation  $F_+(A)=0$ and the Dirac equation $D_A\,q=0$, see e.g.~\cite{Shadchin:2005mx}.   Solutions to the  self-duality equation $F_+(A)=0$ modulo gauge transformations that are trivial at infinity defines the instanton moduli space $\mathfrak{M}^{\text{inst}}$.
The space of solutions to the Dirac equation forms a fiber over the instanton moduli space.

If the gauge group is one of the classical groups $U(N)$, $SO(N)$, $SP(2N)$ (or a product thereof), $\mathfrak{M}^{\text{inst}}$
admits a particular nice description, the ADHM construction  \cite{Atiyah:1978ri}. Let us review this construction for gauge group $U(N)$ and instanton number $k$. 
First introduce the linear data
\begin{equation}
 x\,=\,\left(\mathbf{B}_1,\mathbf{B}_2,\mathbf{I},\mathbf{J}\right)\,\in\,\mathcal{X}_{N,k}\,:=\,
\text{Hom}(V,V)\oplus\text{Hom}(V,V)\oplus\text{Hom}(W,V)\oplus\text{Hom}(V,W)\,,
\label{ADHMmatrices}
\end{equation}
where $V\simeq\mathbb{C}^k$ and $W\simeq\mathbb{C}^N$ carry the defining $GL(k)$ and $GL(N)$ action.
We refer to \eqref{ADHMmatrices} as ADHM matrices. On such matrices the group $GL(2)\times GL(k)\times GL(N)$
acts as follows
\begin{equation}
\left(\mathbf{B}_{\alpha},\mathbf{I},\mathbf{J}\right)\, \mapsto\,
 \left(g_{\phi}\,(M_{\alpha}^{\beta}\,\mathbf{B}_\beta)\,g_{\phi}^{-1},\,g_a\,\mathbf{I}\,g_{\phi}^{-1},\,\det(M)\,g_{\phi}\,\mathbf{J}\,g_{a}^{-1}\right)\,,
\label{GLGLGLaction}
\end{equation}
where $\alpha,\beta=1,2$, the summation over $\beta$ is understood and 
\begin{equation}
M\,\in\,GL(2)\,,\qquad g_{a}\,\in\,GL(N)\,,\qquad g_{\phi}\,\in\,GL(k)\,.
\end{equation}
The next step is to introduce the ADHM equations
\begin{equation}
 \mu_{\mathbb{R}}\,:=\,
\mathbf{I}\,\mathbf{I}^{\dagger}\,-\,\mathbf{J}^{\dagger}\,\mathbf{J}
\,+\,[\mathbf{B}_1,\,\mathbf{B}_1^{\dagger}]\,+\,[\mathbf{B}_2,\,\mathbf{B}_2^{\dagger}]\,{=}\,0\,,
\label{ADHNreal}
\end{equation}
\begin{equation}
 \mu_{\mathbb{C}}\,:=\,\mathbf{I}\,\mathbf{J}+[\mathbf{B}_1,\,\mathbf{B}_2]\,{=}\,0\,.
\label{ADHMcomplex}
\end{equation}
These equations are covariant under the  group action given by \eqref{GLGLGLaction}.
 The $k$-instanton moduli space is decribed as
\begin{equation}
 \mathfrak{M}_{N,k}\,=\,\left\{x\,\in\,\mathcal{X}_{N,k}\,\,\text{such that}\,\,\,\mu_{\mathbb{C}}(x)=0\right\}/GL(k)\,.
\label{modulispace}
\end{equation}
There are two non-trivial steps involved in this definition: (1) impose the ADHM equations, (2) divide the resulting space by the action of $GL(k)$. In the following we will see how these two steps give rise to the contour integral form of the instanton partition function following two different procedures. 
We anticipate that in both cases the integration over $(\phi_1,\dots,\phi_k)$ in \eqref{Zinst} originates from quotienting over $GL(k)$.

\subsection{D-instanton action and its localization}

There is a natural way to understand the ADHM construction  within the language of branes \cite{Witten:1995gx, Douglas:1996uz, Douglas:1995bn}.
In the simplest example, the starting point is a system of $k$ D(-1) branes and $N$ D3-branes .
The idea is to look at this system either as the theory leaving on the  D3-branes, which corresponds to a four dimensional $\mathcal{N}=2$ gauge theory, or as the $d=0+0$ dimensional theory on the D(-1) brane. 
In the latter case one identifies the ADHM equations with D and F-flatness conditions of the auxiliary 0-dimensional theory.
In this way the instanton moduli space $\mathfrak{M}_{N,k}$ is identified with the  Higgs branch of the theory on the D-instanton.

From the point of view of the four dimensional gauge theory the D-instanton  partition function originates as an integral over the k-instanton moduli space.
Following \cite{Moore:1997dj}, see also \cite{Shadchin:2005mx}, one can   present the integration over the instanton moduli space  $\mathfrak{M}_{N,k}$, see \eqref{modulispace}, in terms of the one over $\mathcal{X}_{N,k}$, see \eqref{ADHMmatrices}, and certain ``auxiliary fields''. The latter, as emphasized after  equation \eqref{modulispace}, are introduced in order to implement the ADHM constraints and the operation of taking quotient by $GL(k)$. 
See also \cite{Bruzzo:2002xf, Fucito:2004gi}.

The correspoding partition function reads
\begin{equation}
\mathcal{Z}_k\,=\,\int\,\frac{\mathcal{D}\phi}{\text{Vol}(U(k))}\,
\left[\mathcal{D}x\mathcal{D}\psi\right]
\left[\mathcal{D}\chi\mathcal{D}H\right]
\left[\mathcal{D}\overline{\phi}\mathcal{D}\eta\right]\,
e^{-S_{\Omega}}\,.
\label{instantonACTION}
\end{equation}
This equation needs some explanation. Measure factors in $[\dots]$ correspond to BRST doublets 
\begin{equation}
 \mathcal{Q}\,
\begin{pmatrix}
 x \\
\chi\\
\overline{\phi}
\end{pmatrix}
\,=\,
\begin{pmatrix}
 \psi \\
H\\
\eta
\end{pmatrix}\,,
\qquad
 \mathcal{Q}\,
\begin{pmatrix}
 \psi \\
H\\
\eta
\end{pmatrix}\,=\,
\delta^{\text{torus}}_{\phi,a,\epsilon_{1,2}}\,
\begin{pmatrix}
 x \\
\chi\\
\overline{\phi}
\end{pmatrix}\,.
\end{equation}
The variable $x$ corresponds to ADHM matrixes \eqref{ADHMmatrices}, $\chi=(\chi_{\mathbb{R}},\chi_{\mathbb{C}})$
belong to the same space as the ADHM equations \eqref{ADHNreal}, \eqref{ADHMcomplex} and $\overline{\phi}\,\in\,\text{Hom}(V,V)$.
Their  BRST partners $(\psi,H,\eta)$ have the same transformation properties under $GL(N)\times GL(k)\times GL(2)$.
The variable $\phi$ has no BRST partner, it satisfies $\mathcal{Q}\phi=0$.
$(x,H,\overline{\phi},\phi)$ are bosonic, $(\psi,\chi,\eta)$ fermionic.
The action of $\delta^{\text{torus}}\,\in\,\text{Lie}(\text{torus})$ is defined as taking the infinitesimal version of the transformation   \eqref{GLGLGLaction}  with 
\begin{equation}
 g_a-1\,\sim\,\text{diag}(a_1,\dots,a_N)\,,
\qquad
 g_{\phi}-1\,\sim\,\text{diag}(\phi_1,\dots,\phi_k)\,,
\qquad
 M-1\,\sim\,\text{diag}(\epsilon_1,\epsilon_2)\,.
\label{torusinf}
\end{equation}
It is clear from the action of $\mathcal{Q}$ that $\mathcal{Q}^2\,=\,\delta^{\text{torus}}$. 
The action has the schematic form 
\begin{equation}
 S_{\Omega}\,=\,
\mathcal{Q}\left( \chi\cdot\mu(x)\,+\,t\,\chi\cdot H\,+\,t'\eta[\phi,\overline{\phi}]\,+\,t''\,\psi\,\overline{\phi}\, x\right)\,,
\end{equation}
where $\mu(x)$ are the ADHM equations. The parameter $t,t',t''$ have a fixed value in the original problem, but, as the partition function is independent of their values, we can calculate it for the most convenient choice.  The suffix $\Omega$ refers to the so-called $\Omega$-deformation introduced to regularize the infinite volume of moduli space \cite{Moore:1997dj,Losev:1997bz}.

The variables $(\eta,\chi,H,\overline{\phi})$ come into a quartet that can be integrated out in the following way \cite{Moore:1997dj, Moore:1998et, Shadchin:2005mx}.
Add to the action a Q-exact term
\begin{equation}
 \delta S\,=\,s\,\mathcal{Q}\left(\chi_{\mathbb{R}}\cdot \overline{\phi}\right)\,+\,s'\,\mathcal{Q}\left(x\,\psi\right)\,.
\end{equation}
For $s\rightarrow \infty$, the $H_{\mathbb{R}},\chi_{\mathbb{R}}$ integration  produces the constraints $\overline{\phi}=0=\eta$.
Next taking $t\rightarrow \infty$ the $H_{\mathbb{C}}$ integration is Gaussian and produces a trivial factor, the $\chi_{\mathbb{C}}$ integration is Gaussian (fermionic) and produces the factor 
\begin{equation}
 {\rm det}_{\mu_{\mathbb{C}}}\,\mathcal{Q}^2\,=\,\epsilon^k\,\prod_{1\leq I<J\leq k}\,\left(\phi_{IJ}^2-\epsilon^2\right)\,.
\label{detmu}
\end{equation}
The notation $ \det_{\mu_{\mathbb{C}}}$ refers to taking the determinant in the vector space to which $\chi_{\mathbb{C}}$
belongs which is  the same as the space of ADHM equations $\mu_{\mathbb{C}}$, see \eqref{ADHMcomplex}.
To calculate the determinant we also reduce the $\phi$ integration from $GL(k)$ to its maximal torus
\begin{equation}
 \mathcal{D}\phi\rightarrow\frac{1}{k!}\,\prod_{I=1}^k\,\frac{d\phi_I}{2\pi i}\,\prod_{1\leq I<J\leq k}\,\phi_{IJ}^2\,.
\label{Vander}
\end{equation}
Finally the integration over $(x,\psi)$, is done sending $s'\rightarrow \infty$. The  $(\mathbf{B}_{1,2},\psi_{B_{1,2}})$ integration produces the factor
\begin{equation}
 \frac{1}{\det_{\mathbf{B}_{1,2}}\,\mathcal{Q}^2}\,=\,
\frac{1}{\epsilon_{1,2}^k}\,\prod_{1\leq I<J\leq k}\,\frac{1}{\phi_{IJ}^2-\epsilon_{1,2}^2}\,,
\label{detB}
\end{equation}
and the $(\mathbf{I},,\psi_I)$, $(\mathbf{J},\psi_J)$ integration produces the factor
  \begin{equation}
 \frac{1}{\det_{\mathbf{I},\mathbf{J}}\,\mathcal{Q}^2}\,=\,
\prod_{I=1}^k\,\prod_{\lambda=1}^N\,\frac{1}{\left(\phi_I-a_{\lambda}\right)\left(\phi_I-a_{\lambda}+\epsilon\right)}\,.
\label{detIJ}
\end{equation}
We have thus performed all the integration in \eqref{instantonACTION} except for $(\phi_1,\dots,\phi_k)$.
Putting together \eqref{detmu}, \eqref{Vander}, \eqref{detB}, \eqref{detIJ} one recognizes that $\mathcal{Z}_k$
given in \eqref{instantonACTION} coincides with  \eqref{Zinst}.
This ends the discussion of pure SYM with gauge group $U(N)$.
The discussion can be generalized to other classical gauge groups and quiver gauge theories, see e.g.~\cite{Billo:2012st} and references therein.

\subsection{The instanton partition function from 5d perspective}

As a different picture, we consider the lift of the given $\mathcal{N}=2$ four dimensional gauge theory to five dimensions.
The $\Omega$ deformation can be understood as imposing the identification $(x,y)\sim (e^{\beta\Omega}x,y+\beta)$, where $y$ is the coordinate of the circle of compactification and $\Omega$ the generator of an infinitesimal $SO(4)$ rotation.
 The important fact is that the 5d partition function $\mathcal{Z}_k^{d=5}$ reduces to the twisted Witten index for the supersymmetric quantum mechanics on the instanton moduli space $\mathfrak{M}_{N,k}$ \cite{Nekrasov:1996cz}, as fermions have periodic boundary condition (up to the twist) along $y$ direction.

As descibed in details in \cite{Nekrasov:2004vw}, by the ADHM description of the instanton moduli space \eqref{modulispace},
the 5d index can be given as
\begin{equation}
 \mathcal{Z}_{k}^{d=5}\,=\,
\frac{1}{|W_k|\,\text{Vol}(\mathbb{T}_k)}\,\int_{\mathbb{T}_k}\,
\prod_{I=1}^{k}d\phi_I\,\prod_{\alpha\,\in\,\Delta_+}\left(e^{i\,\langle \alpha, \phi\rangle}-e^{-i\,\langle \alpha, \phi\rangle}\right)\,\frac{\prod(1-\varphi_{\text{ADHM equations}})}{\prod(1-\omega_{\text{ADHM matrices}})}\,.
\label{ABCD}
\end{equation}
The factors $\varphi_{\text{ADHM equations}}$ and  $\omega_{\text{ADHM matrices}}$ are the weights  under  a certain  torus action, 
of which \eqref{torusinf} is the infinitesimal version, of the ADHM equations \eqref{ADHMcomplex} and ADHM matrices \eqref{ADHMmatrices} respectively.
The structure of \eqref{ABCD} originates from the two step procedure emphasized after equation \eqref{modulispace}: the term 
$\prod(1-\varphi_{\text{ADHM equations}})$ encodes the constraints from the ADHM equation, the integration, and the Vandermonde factor, implements the $GL(k)$ quotient.  
In the following we spell out this formula for pure $\mathcal{N}=2$ SYM with gauge group $U(N)$. 

Consider an element in  the maximal torus $\mathbb{T}\,\subset\,GL(k)\times GL(N)\times GL(2)$ given by 
$g_{\phi}\,=\,\text{diag}(e^{i\,\beta\phi_1},\dots,e^{i\,\beta\phi_k})$, $g_{a}\,=\,\text{diag}(e^{i\,\beta a_1},\dots,e^{i\,\beta a_N})$, $M\,=\,\text{diag}(q_1,q_2)$ with $|q_1|,\, |q_2|<1 $,  compare to \eqref{GLGLGLaction}.
The eigenvalues of the  torus action on the ADHM matrices
$\mathbf{B}_{\alpha}$,  $\mathbf{I}$, $\mathbf{J}$ are 
\begin{equation}
  q_{\alpha}\,e^{i\,\beta\,\phi_{IJ}}\,,
\qquad e^{i\,\beta\,(a_{\lambda}-\phi_{I})}\,,
\qquad q_1\,q_{2}\,e^{i\,\beta\,(\phi_{I}-a_{\lambda})}\,,
\end{equation}
where $\alpha=1,2$, $I,J=1,\dots,k$, and $\lambda=1,\dots,N$. One thus obtains 
\begin{equation}
 \frac{1}{\prod\,(1-\omega_{\text{ADHM matrices}})}\Big{|}_{\mathbf{I},\mathbf{J}}\,=\,
\prod_{I=1}^k\,\frac{1}{P_{+\beta}(e^{-i\,\beta\,\phi_{I}})\,P_{-\beta}(q_1\,q_2\,e^{i\,\beta\,\phi_{I}})}\,,
\end{equation}
\begin{equation}
 \frac{1}{\prod\,(1-\omega_{\text{ADHM matrices}})}\Big{|}_{\mathbf{B}_{1,2}}\,=\,
\frac{1}{(1-q_{1,2})^k}\,\frac{1}{\Delta_{\beta}(q_{1,2})}\,,
\end{equation}
where
\begin{equation}
 P_{\beta}(t)\,:=\,\prod_{\lambda=1}^N\,(1-t\,e^{i\,\beta\,a_{\lambda}})\,,
\qquad 
\Delta_{\beta}(t)\,:=\,\prod_{I\neq J}\,(1-t\,e^{i\,\beta\,\phi_{IJ}})\,.
\end{equation}
It is not a coincidence that these equations are in correspondence with \eqref{detIJ}, \eqref{detB}.

Next consider the the contribution from the \emph{complex} ADHM equations \eqref{ADHMcomplex}.
Its weight under the thorus action gives 
\begin{equation}
 \prod\,(1-\varphi_{\text{ADHM equations}})\,=\,(1-q_1\,q_2)^k\,\Delta_{\beta}(q_1\,q_2)\,.
\end{equation}
Compare to \eqref{detmu}.
Finally the Vandermonde is equal to $\Delta_{\beta}(1)$.
Putting things together according to equation \eqref{ABCD} we obtain
\begin{equation}
\frac{(1-q_1\,q_2)^k}{(1-q_1)^k(1-q_2)^k}\,\frac{\Delta_{\beta}(1)\,\Delta_{\beta}(q_1\,q_2)}{\Delta_{\beta}(q_1)\,\Delta_{\beta}(q_2)}\,
\prod_{I=1}^k\,\frac{1}{P_{\beta}(e^{-i\,\beta\,\phi_{I}})\,P_{\beta}(q_1\,q_2\,e^{-i\,\beta\,\phi_{I}})}\,. 
\end{equation}
The cases of $SO(N)$ and $SP(2N)$ are similar upon modifying ADHM data and related thorus action,  \cite{Nekrasov:2004vw}, see also \cite{Marino:2004cn}.
Matter can also be included in this picture.
The limit $\beta\rightarrow 0$ can be easily taken upon setting $q_{1,2}\,=\,e^{i\,\beta\epsilon_{1,2}}$.
Notice that the condition  $|q_1|,\, |q_2|<1 $ translates into a small imaginary part for $\epsilon_{1,2}$.

\subsection{Building blocks for quiver gauge theories}
\label{app:quiverblocks}

We now present the recipe to obtain the integrand for the instanton partition function corresponding to  the class of quiver gauge theories introduced in Section \ref{sec:quiver}.
To which we refer for the notation used here. 
From the discussion above it is expected that the integrand  should  be written in terms of the following data:
\begin{itemize}
 \item[(1)] an element $q=q(\epsilon_{1,2},a,\phi,m)\,\in\,\mathbb{T}\subset SO(4)_{\text{Lorentz}}\times G_{\text{gauge}}\times  G_{\text{inst}}\times  G_{\text{flavour}}$,
\item[(2)] a ``collection'' of weights $\omega_{\ell}$, more precisely   eigenvalues $e^{i\, \omega_{\ell}}$,   under this torus action.
\end{itemize}
From this data the integrand can be constructed using the following rule
\begin{equation}
 \sum_{\ell}\,n_{\ell}\,e^{i\, \omega_{\ell}}\,\mapsto\,\prod_{\ell}\,(\omega_{\ell})^{-n_\ell}\,,
\label{rulecharacter}
\end{equation}
where $n_{\ell}=\pm 1$.

Let us introduce the following modules for the torus action:
\begin{equation}
 W_{\mathsf{v}}\simeq\mathbb{C}^{N_{\mathsf{v}}}\,,
\qquad
 V_{\mathsf{v}}\simeq\mathbb{C}^{k_{\mathsf{v}}}\,,
\qquad
 M_{\mathsf{v}}\simeq\mathbb{C}^{n_{\mathsf{v}}}\,,
\qquad
L\simeq \mathbb{C}^2\simeq \mathbb{R}^4\,.
\end{equation}
They carry, respectively, the defining representation of  $GL(N_{\mathsf{v}})$, $GL(k_{\mathsf{v}})$, $GL(n_{\mathsf{v}})$,  $SO(4)_{\text{Lorentz}}$  
and thus of the maximal torus. 
We denote the corresponding characters as $\chi_{W,V,M}=\text{tr}_{W,V,M}(q)$, for example $\chi_{V}=\sum_{I=1}^ke^{i\,\phi_I}$.
We denote their conjugate modules as  $\overline{W}_{\mathsf{v}}$, $\overline{V}_{\mathsf{v}}$, $\overline{M}_{\mathsf{v}}$.

From the description of the tangent space to the instanton moduli space in terms of linearized ADHM equations and infinitesimal gauge transformations, see \cite{Nekrasov:2002qd}, one finds, for each gauge group factor 
\begin{equation}
 \text{Ch}(T\mathfrak{M}_{N,k})\,=\,\chi_{\overline{W}}\,\chi_{V}\,-\,(e^{i\,\epsilon_1}-1)(e^{i\,\epsilon_2}-1)\,\chi_{V}\,\chi_{\overline{V}}\,+\,e^{i\,\epsilon}\,\chi_{\overline{V}}\,\chi_{W}\,.
\label{Chgauge}
\end{equation}
Appling the rule \eqref{rulecharacter} to this character one obtains the integrand \eqref{Zinst}.
The contribution of bifundamental matter  is given by
\begin{equation}
 \text{Ch}(\square_\mathsf{v}, \overline{\square}_\mathsf{w})\,=\,-\,e^{i\,m_{\mathsf{v}\,\mathsf{w}}}
\left[\chi_{W_{\mathsf{v}}}\,\chi_{\overline{V}_{\mathsf{w}}}\,-\,(e^{i\,\epsilon_1}-1)(e^{i\,\epsilon_2}-1)\,
\chi_{V_{\mathsf{v}}}\,\chi_{\overline{V}_{\mathsf{w}}}\,+\,e^{i\,\epsilon}\,\chi_{V_{\mathsf{v}}}\,\chi_{\overline{W}_{\mathsf{w}}}\right]\,.
\end{equation}
Notice the overall minus sign compared to \eqref{Chgauge}.
Applying the rule \eqref{rulecharacter}   one  obtains
\begin{equation}
 \text{Ch}(\square_\mathsf{v}, \overline{\square}_\mathsf{w})\,\mapsto\,(-1)^{k_{\mathsf{w}N_{\mathsf{v}}}}\,\prod_{I=1}^{k_{\mathsf{v}}}\prod_{J=1}^{k_{\mathsf{w}}}
\frac{P_{\mathsf{v}}(\phi_{\mathsf{w},J}-m_{\mathsf{v}\,\mathsf{w}})\,P_{\mathsf{w}}(\phi_{\mathsf{v},I}+m_{\mathsf{v}\,\mathsf{w}}+\epsilon)}{\Delta(m_{\mathsf{v}\,\mathsf{w}}+\phi_{\mathsf{v},I}-\phi_{\mathsf{w},J})}\,=:\,\tilde{z}^{\mathsf{v}\,\mathsf{w}}_{k_\mathsf{v}\,k_\mathsf{w}}\,.
\end{equation}
From this expression it follows that the contribution from  $ \text{Ch}(\square_\mathsf{v}, \overline{\square}_\mathsf{w})+ \text{Ch}(\square_\mathsf{w}, \overline{\square}_\mathsf{v})$ is given by
\begin{equation}
 z^{\text{bifund}}_{k}\,=\,\tilde{z}^{\mathsf{v}\,\mathsf{w}}_{k_\mathsf{v}\,k_\mathsf{w}}\,\tilde{z}^{\mathsf{w}\,\mathsf{v}}_{k_\mathsf{w}\,k_\mathsf{v}}\,.
\end{equation}
Finally fundamental matter gives the character
\begin{equation}
 \text{Ch}(\square_\mathsf{v})\,=\,-\chi_{V}\,\chi_{\overline{M}}\ \ \stackrel{\text{\eqref{rulecharacter} }}{\longrightarrow}\ \ z^{\text{fund}}_k\,=\,\prod_{I=1}^{k}\prod_{f=1}^n\left(\phi_I-m_f\right)\,.
\end{equation}
The expressions above can also be found in \cite{Shadchin:2005cc, Billo:2012st}.

\subsection{Contour and classification of poles}
\label{App:contourRES}
The poles contributing to \eqref{Zinst} correspond to solutions of simple algebraic equations. 
As  we will now review following \cite{Litvinov:2013zda}, these solutions are classified by N-tuple of Young diagrams \cite{Nekrasov:2002qd}.
Let 
\begin{equation}
 A(x):=\prod_{\lambda=1}^N(x-a_{\lambda})\,,
\qquad
E(x):= \frac{(x-\epsilon_1)(x-\epsilon_2)}{x(x-\epsilon_1-\epsilon_2)}\,.
\end{equation}
We want to classify solutions of
\begin{equation}
 A(x_i)\,\prod_{j\neq i}E(x_i-x_j)\,,\qquad \text{for}\,\,\,\,\,i=1,\dots,k,
\label{equation}
\end{equation}
modulo permutations of $\{x_i\}$. Notice that, compare to \eqref{Zinst}, only ``half of the poles'' of the integrand contributes. This is so as a consequence of the choice of contour.
It is clear that at least one $x$ has to be a zero of $A$. Without loss of generality we set $x_1=a_{\lambda_1}$ and get $k-1$ equations
\begin{equation}
 A'(x_i)\,\prod_{\stackrel{j>1}{j\neq i}}E(x_i-x_j)\,,\qquad \text{for}\,\,\,\,\,i=2,\dots,k,
\label{reduced}
\end{equation}
where $A'(x):=A(x)E(x-a_{\lambda_1})$.
For the same argument as  above applied now to \eqref{reduced}, we  set $A'(x_2)=0$.
We have to recall that we must keep $x_i-x_j\neq 0$  and $x_i-x_j\neq \epsilon_1+\epsilon_2$ otherwise the denominator in $E(x)$ vanishes.  
Iterating this procedure one finds that solutions of \eqref{equation} modulo permutations are classified by the N-tuple of Young diagrams.
For example for $k=3$ one has
\begin{equation}
 \{a_{\lambda_1},a_{\lambda_2},a_{\lambda_3}\}\,,
\,\,\,
\{a_{\lambda_1},a_{\lambda_2},a_{\lambda_2}+\epsilon_{\alpha}\}\,,
\,\,\,
\{a_{\lambda_1},a_{\lambda_1}+\epsilon_{\alpha},a_{\lambda_1}+2\epsilon_{\alpha}\}\,,
\,\,\,
\{a_{\lambda_1},a_{\lambda_1}+\epsilon_1,a_{\lambda_1}+\epsilon_{2}\}\,,
\end{equation}
where $\alpha=1,2$ and the $a_{\lambda}$ are different from each other. 


\section{Useful formulas}
\label{App:UsefulFormula}

\subsection{Computing \eqref{ZD=0res}}

In the following we present the explicit calculation of the residue corresponding to the pole \eqref{poleF=0}.
First we introduce the notation
\begin{equation}
 \mathbf{Y}_1\,=\,\{1,\dots,s_1\}\,,
\quad
 \mathbf{Y}_2\,=\,\{s_1,\dots,s_1\,+\,s_2\}\,,
\,\,\,
\dots
\,\,\,,
 \mathbf{Y}_N\,=\,\{s_1+\dots+s_{N-1},\dots,k\}\,.
\qquad
\end{equation}
The next step is to split the interaction term $\prod\,\frac{\phi_{IJ}^2}{\phi_{IJ}^2-\epsilon_2^2}$ in the integrand as
\begin{equation}
\prod_{1\,\leq I<J\leq k}\frac{\phi_{IJ}^2}{\phi_{IJ}^2-\epsilon_2^2}
\,=\,  
\left[\prod_{\lambda=1}^N\,
\prod_{\substack{I<J\\I,J\,\in\,\mathbf{Y}_{\lambda}}}\frac{\phi_{IJ}^2}{\phi_{IJ}^2-\epsilon_2^2}\right]\,
\prod_{\lambda < \lambda'}\prod_{I_\lambda, I_{\lambda'}}\frac{\phi_{I_{\lambda}I_{\lambda'}}^2}{\phi_{I_{\lambda}I_{\lambda'}}^2-\epsilon_2^2}\,,
\end{equation}
Only the factor in square bracket on the right hand side is singular in the limit in which $\phi_I$ take the value give by \eqref{poleF=0}. For these terms we use the simple result
\begin{equation}
 \lim_{\phi_{\kappa_{\lambda}+I_{\lambda}}\rightarrow a_{\lambda}+(I_{\lambda}-1)\epsilon_2}\,
\prod_{I=2}^{s_{\lambda}}\left(\phi_{\kappa_{\lambda}+I_{\lambda}}-a_{\lambda}-(I_{\lambda}-1)\epsilon_2\right)\,
\prod_{\substack{I<J\\I,J\,\in\,\mathbf{Y}_{\lambda}}}\,\frac{\phi_{IJ}^2}{\phi_{IJ}^2-\epsilon_2^2}\,=\,
\frac{\epsilon_2^{{s_{\lambda}}-1}}{s_{\lambda}}\,,
\end{equation}
We also use
\begin{equation}
 \lim_{\phi_{\kappa_{\lambda}+1}\rightarrow a_{\lambda}}\,
\left(\phi_{\kappa_{\lambda}+1}-a_{\lambda}\right)\,\prod_{I\,\in\,\mathbf{Y}_{\lambda}}\,Q(\phi_I)\,=\,
\frac{1}{\epsilon_2^{s_{\lambda}-1}}\,\frac{1}{(s_{\lambda}-1)!}\,
\prod_{I=1}^{s_{\lambda}}\,\frac{Q^{(\lambda)}(a_{\lambda}+(I-1)\epsilon_2)}{\epsilon_1+I\epsilon_2}\,.
\end{equation}
where
\begin{equation}
 Q^{(\lambda_1)}(x)\,=\,\frac{1}{P^{(\lambda_1)}(x)P^{(\lambda_1)}(x+\epsilon)}\,,\qquad
P^{(\lambda_1)}(x)\,=\,\prod_{\lambda\neq\lambda_1}(x-\lambda)\,.
\end{equation}
The result \eqref{ZD=0res} immediately follows.

\subsection{Exponentiation of residues}

One has
\be 
\sum_{k=0}^\infty {q^k \over k! \epsilon^k}\, \prod_{I=1}^k f(x+(I-1)\epsilon)\, =\,
 \exp \left( {1\over \epsilon} \,
\sum_{\ell=1}^\infty
 {q^\ell \over \ell^2}\left(\frac{1}{(\ell-1)!} {d^{\ell-1} \over d x^{\ell-1}}f(x)^\ell \right) + {\cal O}(\epsilon^0) \right)
\label{formulaforres} \ee
as  formal series in $\epsilon$.

We present the following generalization of \eqref{formulaforres}
\bea
&& \hskip -1cm \sum_{s_1,\dots,s_{N}\geq 1}\,
\prod_{\lambda=1}^N\,
\left(
\frac{q^{s_{\lambda}}\, \prod_{I=1}^{s_\lambda} A(x_\lambda+(I-1)\epsilon))}{s_{\lambda}!\,\epsilon^{s_{\lambda}}}\right)\,
\prod_{1\leq\lambda < \lambda'\leq N} \, \prod_{I_\lambda=1}^{s_\lambda} \, \prod_{I_{\lambda'}=1}^{s_{\lambda'}} {(x_{\lambda \lambda'} + (I_\lambda - I_{\lambda'}) \epsilon)^2 \over (x_{\lambda \lambda'} + (I_\lambda - I_{\lambda'}) \epsilon)^2 - \epsilon^2} \nonumber
\\
&& \qquad \,=\, \exp \left( {1\over \epsilon} \,
\sum_{\lambda=1}^N\,\sum_{\ell=1}^\infty
 {q^\ell \over \ell^2}\left(\frac{1}{(\ell-1)!} {d^{\ell-1} \over d x_\lambda^{\ell-1}}A(x_\lambda)^\ell \right) 
\,+\, {\cal O}(\epsilon^0) \right)
\label{usufulresidue2}
\eea
as  formal series in $\epsilon$.
We do not present a proof of this result, but it can be easily checked order by order in $q$.

\subsection{Illustrative Examples for Iterated Mayer expansion}
\label{App:ExampleIteratedMayer}

In the following we spell out the definition \eqref{Sgothicdef} for $k=2,3$.
For $k=2$ \eqref{Sgothicdef} reads
\begin{equation}
 \mathfrak{S}_{\{\{1,2\}\}}=\,S_{\{1,2\}}\,\mathfrak{g}_{\{1,2\}}\,=\,\left(\mathcal{D}_{12}-1\right)\,
\widetilde{\mathcal{D}}_{12}\,,
\qquad
 \mathfrak{S}_{\{\{1\},\{2\}\}}=\,\mathfrak{f}_{\{1\},\{2\}}\,=\,\widetilde{\mathcal{D}}_{12}-1\,,
\end{equation}
and the sum of this two terms is by the definitions above $\mathcal{F}_{12}$.
For $k=3$, \eqref{Sgothicdef} reads
\bea
  \mathfrak{S}_{\{\{1,2,3\}\}} &= & S_{\{1,2,3\}}\,\mathfrak{g}_{\{1,2,3\}}\,,
\\
  \mathfrak{S}_{\{\{I,J\},\{K\}\}} &= & S_{\{I,J\}}\,\mathfrak{g}_{\{I,J\}}\,\mathfrak{f}_{\{\{I,J\},\{K\}\}}\,,
\\
  \mathfrak{S}_{\{\{1\},\{2\},\{3\}\}} &= & \mathfrak{f}_{\{\{1\},\{2\}\}}\,\mathfrak{f}_{\{\{2\},\{3\}\}}+
\mathfrak{f}_{\{\{1\},\{3\}\}}\,\mathfrak{f}_{\{\{2\},\{3\}\}}+
\mathfrak{f}_{\{\{1\},\{2\}\}}\,\mathfrak{f}_{\{\{1\},\{3\}\}} \nonumber \\
  &&+
\mathfrak{f}_{\{\{1\},\{2\}\}}\,\mathfrak{f}_{\{\{2\},\{3\}\}}\,\mathfrak{f}_{\{\{1\},\{3\}\}}\,.
\eea

\section{Expansion by regions for instanton partition functions}
\label{app:region}

One recurrent question in this paper is how to evaluate multiple integrals as a Laurent expansion in some
 small parameter\footnote{We are actually interested only in the leading $\frac{1}{\epsilon_2}$ behavior of the logarithm of the partition function.}, namely $\epsilon_2$.
There is a powerful method to address this question, it is the 
so called method of \emph{expansion by regions}, usually applied to the evaluation of Feynman integrals  \cite{Beneke:1997zp, {Smirnov:2002pj}}. 
The method can briefly summarized as follows: 
(1) divide the integration domain into regions and expand the integrand in a Taylor series in small parameters,
(2) extend the integration to the full domain of integration,
(3) set the scaleless integrals  to zero. 
It has been applied successfully in many situations in the context of Feynman integrals, see \cite{Jantzen} for a recent study including proof under some assumptions. 
We apply the same idea to the integrals of instanton partition functions.

It might look surprising that the prescription above works at all. In particular one may worry about multiple counting of contributions.
Let us describe how this problem is circumvented in a simple example, the $k=2$ case in \eqref{ZwithF=0}
\be F = \,\int   \,  {d \phi_1 \over 2\pi i} {d \phi_2 \over 2\pi i}\,{ Q(\phi_1)}{ Q(\phi_2)} \, {\phi_{12}^2 \over \phi_{12}^2 - \epsilon_2^2} \,.
\label{ZwithF=0_k=2} \ee
The first step is to identify the set of relevant regions. 
For this, one may note that the interaction term ${\phi_{12}^2 \over \phi_{12}^2 - \epsilon_2^2}$ 
is different from one only in the \emph{region} in which $\phi_{12}$ is of order $\epsilon_2$.
We thus define two separate integration regions as follows
\begin{equation}
D^{\bf S} \ : \ \ |\phi_{12}| \,\sim\,{\cal O}(\epsilon_2) \,,  \qquad {\rm and} \qquad
D^{\bf L} \ : \ \ |\phi_{12}| \, \gg\,\epsilon_2 \,,
\end{equation}
where ${\bf S}$ and ${\bf L}$ denote  short and long distance respectively, and the whole integration domain is given as $D= D^{\bf S} \cup D^{\bf L}$. 

The next step is to Taylor expand the integrand in each region. To do so it is convenient to perform the change of variables 
$\phi_0 := {\phi_1 + \phi_2 \over 2}$, $\tilde\phi := {\phi_1 - \phi_2}$. 
%
%
%
%
In the  region $D_{\bf S}$ one Taylor-expands the $Q$ functions for small\,\footnote{Note that $\phi_0$ will be equal to the poles given by Coulomb parameters $a_\lambda \gg \epsilon_2$.}
 $\tilde\phi$, and in region $D_{\bf L}$ one expands the interaction term ${\tilde\phi^2 \over \tilde\phi^2 - \epsilon_2^2}$. 
Accordingly we  introduce the Taylor expansion operators as follows
\bea
\mathbb{T}^{\bf S} [I] &=& \sum_{k=0}^\infty \mathbb{T}^{\bf S}_k [I] \,=\, \sum_{k=0}^\infty  \, {\cal Q}_k(\phi_0)\,\tilde\phi^{2k} \,
\,{ \tilde\phi^2 \over \tilde\phi^2-\epsilon_2^2} \,, \label{TSk2} \\
\mathbb{T}^{\bf L}[I] &=& \sum_{l=0}^\infty \mathbb{T}^{\bf L}_l [I] \,=\, \sum_{l=0}^\infty   Q(\phi_1)Q(\phi_2)\, \Big({\epsilon_2^2 \over\tilde\phi^2} \Big)^l\,, \label{TLk2}
\eea
where
\begin{equation}
 {\cal Q}_k(x)\,:=\,\frac{1}{2^{2k}}\,\sum_{m=-k}^{k}\,\frac{(-1)^{k+m}}{(k+m)!(k-m)!}\,Q^{(k-m)}(x)\,Q^{(k+m)}(x)\,. \label{calQ}
\end{equation}

After the Taylor expansion, one extends the integration region to be the full region $D$. According to the prescription, one has
\be F = F^{\bf S} + F^{\bf L} \,, 
%
%
\qquad
F^{\bf S} = \sum_{k=0}^\infty \int_D \mathbb{T}^{\bf S}_k [I]  \,, \qquad
F^{\bf L} = \sum_{l=0}^\infty \int_D \mathbb{T}^{\bf L}_l [I] \,.
\label{Feq}
\ee
The fact that this equality holds is  a very non-trivial statement.
 In particular one may worry about multiple-counting of contributions. 
One can verify  \eqref{Feq} directly upon giving a certain  prescription for the evaluation of the relevant integrals.
Here we provide a formal argument, following closely \cite{Jantzen}, which can be  easily  generalized to the $k$-tuple integrals entering the instanton partition function. 
An honest decomposition of \eqref{ZwithF=0_k=2} will proceed as follows
\be F 
=  \sum_k \int_{D^{\bf S}} \mathbb{T}^{\bf S}_k [I] +
\sum_l \int_{D^{\bf L}} \mathbb{T}^{\bf L}_l [I] 
= F^{\bf S} + F^{\bf L}  - \Big( \sum_k \int_{D^{\bf L}} \mathbb{T}^{\bf S}_k [I] +
\sum_l \int_{D^{\bf S}} \mathbb{T}^{\bf L}_l [I] \Big) \,. \label{honestdec} \ee
%
%
In order to show that this is equal to  \eqref{Feq}, one has to show that the terms in the bracket give zero. 
These terms can be written nicely as
\be \sum_k \int_{D^{\bf L}} \sum_l \mathbb{T}^{\bf L}_l \big[ \mathbb{T}^{\bf S}_k [I] \big] +
\sum_l \int_{D^{\bf S}} \sum_k \mathbb{T}^{\bf S}_k \big[ \mathbb{T}^{\bf L}_l [I] \big] = \sum_{k,l} \int_D \mathbb{T}^{(\bf S, L)}_{j,l} [ I ]  \,, \label{TLSij} \ee
where we use the fact that the double expansion of the integrand is independent of the  order in which short or long range expansions are applied\,\footnote{This corresponds to the case of ``commuting expansion" in \cite{Jantzen}.}, so that
\be\mathbb{T}^{\bf L}_l \, \mathbb{T}^{\bf S}_k = \mathbb{T}^{\bf S}_k \, \mathbb{T}^{\bf L}_l = : \mathbb{T}^{(\bf S, L)}_{k,l} \,. \label{doubleTAYLOR}\ee
Explicitly, each term in \eqref{TLSij} is given as

\be \int_D \mathbb{T}^{(\bf S, L)}_{j,l} [ I ]  \,=\,  
\epsilon_2^{2l}\, \int {d\phi_0 \over 2 \pi i }\,
{\cal Q}_k(\phi_0)\,{d\tilde\phi \over 2 \pi i }\, (\tilde\phi^2)^{k-l} \,. \ee
This is indeed zero since
\be \int d\tilde\phi \, (\tilde\phi^2)^n = 0 \,, \qquad {\rm for~integer~} n \,. \label{scaleless} \ee
One can take \eqref{scaleless} as the analogue of setting ``scaleless integrals" to  zero in the case of Feynman integrals. 
As in that case, this step is essential in order for the method of expansion by regions to  work.

The above considerations seem to be easily generalized to the multi-particle cases. In this case one deals with $k$-folded integrals.
It is expected that the double Taylor expansion in different regions, compare to \eqref{doubleTAYLOR}, will produce some factorized term in the form of $\int d\tilde\phi \, (\tilde\phi)^n$.
This is set to zero as in \eqref{scaleless}.

Before proceeding to some concrete examples, let us make some important remarks.
When applied to the calculation of Feynman integrals,  the method of expansion by regions presents the characteristic  feature of introducing extra divergences in individual integrals.
These new divergences must be regularized and should cancel after summing over all regions in order to reproduce the correct result\,\footnote{For example, in the dimensional regularization $D=4-2\varepsilon$, this corresponds to higher order ${1\over \varepsilon}$ singularities, corresponding to either UV or IR divergence. The new UV and IR divergences cancel with each other from different regions. These divergence may also be understood as from the scaleless integral, where the UV and IR divergences canceled within the scaleless integral itself (see examples in \cite{Jantzen}).}. The situation in the case of the integrals entering the instanton partition function is somehow similar.
By staring at \eqref{TSk2}-\eqref{TLk2} one realizes that upon Taylor expansion, one produces arbitrary integer powers of  $\tilde\phi^2$. These are positive powers for  $\mathbb{T}^{\bf S}[I]$ and negative for $\mathbb{T}^{\bf L}[I]$. The integration along the real line would produce  higher order UV divergence in the short regions expansion, and higher order IR divergence in the long range expansion.
As in the case of Feynman integrals these divergences needs to be regularized.
In the examples below the regularization corresponds to a certain contour integral interpretation of the integrals.
This prescription works in all examples we considered. We recall that the leading term in $\epsilon_2$, the term we are actually interested in, does not need any regularization of this type.


\subsection{Checks of expansion by regions}

In this appendix we check the method of expansion by regions via explicit residue computations for two cases. For simplicity, we will consider the $U(1)$ case, namely in the following we take 
\be
Q(x)={f(x) \over x - a -i\,0}\,, \qquad f(x) : = {1\over x -a +\epsilon + i 0}\,,
\ee
where we introduce $f(x)$ for convenience as it does not have any pole inside the contour of integration.

\subsubsection*{Example one}

We first consider the integral \eqref{ZwithF=0_k=2}. It is straightforward to compute it by summing over the residues of two poles: $(\phi_1, \phi_2)=(a,a+\epsilon_2), (a+\epsilon_2,a)$. One has
\be F = {f(a) f(a+\epsilon_2) } = \sum_{k=0}{\epsilon_2^{k} f(a) f^{(k)}(a) \over  k!} \,. \label{residueF} \ee
On the other hand, by the method of expansion by regions, the integral is also given as
the sum $F_{\bf S}+F_{\bf L}$.
One can check that indeed this form reproduces the residue result order by order in $\epsilon_2$ expansion. 
It is interesting that $F_{\bf S}$  contributes only to the terms of odd power of $\epsilon_2$, while $F_{\bf L}$ contributes to the terms of even power.

Let us spend a few more words on the integrals in $F_{\bf L}$
\be
\int {d \phi_1 \over 2\pi i} {d \phi_2\over 2\pi i} {f(\phi_1)f(\phi_2) \over (\phi_1 -a-i0) (\phi_2-a-i0) (\phi_1 - \phi_2)^{2k}} \label{integralAPPEx1}\,.
\ee
The main issue is how to treat the new denominators $\phi_{12}^{2k}$ in the expansion. We will apply the following prescription. We first do integration for $\phi_1$. There is a simple pole at $\phi_1=a + i0$. After taking the residue, the remaining integral for $\phi_2$ has a $(2k+1)$th-order pole of $\phi_2 = a$, the finally residue is given as
\be { f(a) f^{(2k)}(a) \over 2k!} \,. \label{FLresidue} \ee
This prescription works for the next example as well. This provides a strong support for it.

\subsubsection*{Example two}

We consider an integral which appear at $k=3$ order in \eqref{LogZF=0}
\be F= \int \prod_{I=1}^3 {d\phi_I \over 2\pi i} {Q(\phi_I)} \, {1 \over \phi_{12}^2 - \epsilon_2^2} {1 \over \phi_{13}^2 - \epsilon_2^2}  \, . \label{ApptoCOMUPUTE}\ee
As in the  example above, one can compute the integral directly by residues. As there is no $\phi_{IJ}$ terms in the numerators,
also poles for which different $\phi$'s are at the same position contribute.
Let us summarize the result of the residue calculation. 
The poles of $(\phi_1, \phi_2, \phi_3)$ and the corresponding residues are given  as follows:
\bea
(a,a,a)\,: &\quad& {f(a)^3 \over \epsilon^4} \,, \\
(a,a,a+\epsilon_2)+(a,a+\epsilon_2,a)\,: &\quad& -{f(a)^2 f(a+\epsilon_2) \over \epsilon^4} \,, \\
(a+\epsilon_2,a,a)\, \footnotemark\,: &\quad& {f(a)^2 f'(a+\epsilon_2) \over 4 \epsilon^3} - {f(a)^2 f(a+\epsilon_2) \over 2 \epsilon^4} \,, \\
(a,a+\epsilon_2,a+\epsilon_2)\,: &\quad& {f(a) f(a+\epsilon_2)^2 \over 4 \epsilon^4} \,, \\
(a+\epsilon_2,a+\epsilon_2,a) + (a+\epsilon_2,a, a+\epsilon_2)\,: &\quad& {f(a)f(a+\epsilon_2)f(a+2\epsilon_2) \over 4\epsilon^4} \,.
\eea
\footnotetext{Note in this case, $\phi_1=a+\epsilon_2$ should be taken as a double pole, due to $\phi_{12} -\epsilon_2$ and $\phi_{13}-\epsilon_2$ in the denominator.}%
One can see that each pole starts at the order $1/\epsilon_2^4$. However, summing over the residues one finds that the first two orders are cancelled, the leading order starts at $1/\epsilon_2^2$, which gives
\be {6 f(a) f'(a)^2 + 3 f(a)^2 f''(a) \over 8 \epsilon_2^2} + {3 f(a) f'(a) f''(a) + f(a)^2 f^{(3)}(a)\over 3\epsilon_2}  + {\cal O}(\epsilon_2^0) \,. \label{residueleading2} \ee
While for our applications, see \eqref{LogZF=0},  we are interested only in the leading contributions  (in the normalization used here this is at order $\frac{1}{\epsilon_2^2}$). In the following we will verify that also the next to leading contribution is reproduced by the method of expansion by regions.

Let us compute \eqref{ApptoCOMUPUTE} by the method of expansion by regions. According to the discussion in Section \ref{byregions}, there are five regions to consider in this case, we have
\be F = F^{\{\{1,2,3\}\}}  + F^{\{ \{1,2\},3\}} +F^{\{\{1,3\},2\}} +F^{\{\{2,3\},1\}} + F^{\{\{1\},\{2\},\{3\}\}} \,. \label{Finregions}\ee
In the first region, there is only one cluster, and one only Taylor expands the $Q$ functions around the center of cluster coordinate $\overline{\phi}$, where
\be \overline{\phi} : = {\phi_1+\phi_2 + \phi_3 \over 3} \,, \qquad \tilde\phi_I := \phi_I - \phi_0 \, , \ee
and $\sum_{I=1}^3 \tilde\phi_I=0$. One has
\be
F^{\{\{1,2,3\}\}}=
\int d \overline{\phi}\,\prod_{I=1}^3 {d\tilde\phi_I\over 2\pi i}\, \delta\Big({\sum_{I=1}^3 \tilde\phi_I \over 3}\Big)  \left(Q^3(\overline{\phi})+ 0\times \phi_I + \mathcal{O}(\tilde\phi_I^2)\right)\,
 {1 \over \tilde\phi_{12}^2 - \epsilon_2^2} {1 \over \tilde\phi_{13}^2 - \epsilon_2^2}  \,. 
\label{ex2-leading}
\ee
The leading term is straightforward to compute. By integrating out $\tilde\phi_I$'s one has
\be {1\over 4 \epsilon_2^2} \int {d\overline{\phi} \over 2\pi i} Q(\overline{\phi})^3 \,, \ee
which after integrating $\phi_0$ exactly reproduces the leading contribution in \eqref{residueleading2}. 
The remaining  terms in the Taylor expansion will contribute to ${\cal O}(\epsilon_2^0)$ order (since $\tilde\phi_I^2 \sim \epsilon_2^2$).
These will constribute to the $\mathcal{O}(\epsilon_2^0)$ order in \eqref{residueleading2}.

In the next relevant region, there are two clusters. Introduce the center coordinate for the two-particle cluster as
\be \overline{\phi}_a := {\phi_1 + \phi_2 \over 2}\,,\,\,\,\,\,\,\,a\,=\,\{1,2\} \,, \qquad \tilde\phi := \phi_1 -\phi_2 \,. \ee
and set the coordinate in the one particle cluster $\overline{\phi}_b=\phi_3$, where $b=\{3\}$.
Doing Taylor expansion one has
\be
F^{\{\{1,2\},\{3\}\}}\,=\,
\int {d\overline{\phi}_a\over 2\pi i}\,{d\overline{\phi}_b\over 2\pi i}\,{d\tilde\phi \over 2\pi i}
\, { Q^2(\overline{\phi}_a)Q(\overline{\phi}_b) + {\cal O}(\tilde\phi, \epsilon_2^2)
  \over \overline{\phi}_{ab}^2\,(\tilde\phi^2 - \epsilon_2^2)}
\ee
where ${\cal O}(\tilde\phi, \epsilon_2^2)$ corresponds to higher order terms  in the Taylor expansion, which are subleading in the small $\epsilon_2$ expansion.
To compute the leading term in this region, one needs to evaluate the integral
\be 
\int {d x \over 2\pi i} {d y \over 2\pi i} {f(x)^2 f(y) \over (x -a-i0)^2 (y-a-i0) (x - y)^2} \,,
\ee
where, for simplicity, we considered the case corresponding to the $U(1)$ gauge theory.
This integral is similar to \eqref{integralAPPEx1}.
To evaluate it,  one can first integrate out $x$ by taking residue at the double pole $x=a+i0$ then do integration for $y$. One can also first do integration for $y$ and then for $x$, the result is the same. 
By summing the leading contributions from both regions\footnote{The region $\{\{2,3\},\{1\}\}$ in this case contributes to higher order, since there is no ${1\over \phi_{23}^2 -\epsilon_2^2}$ factor in the integral.} $\{\{1,2\},\{3\}\}$ and $\{\{1,3\}, \{2\}\}$, 
one reproduces the next-to-leading order term in \eqref{residueleading2}.
It is clear that  the region ${\{\{1\},\{2\},\{3\}\}}$  where all particles are well separated, compare to \eqref{Finregions}, contributes  ${\cal O}(\epsilon_2^0)$ to \eqref{residueleading2}. 

This example provides a non-trivial check for the correctness of the method of expansion by regions. In particular, it shows that if one is only interested in the leading contribution, 
the method of expansion by regions turns out to be extremely efficient.


\section{Alternative derivation of \eqref{mainresultfromMayer}}
\label{App:regionsplusMayer}

In this Appendix we rederive \eqref{mainresultfromMayer}  by  applying the method of expansion by regions before 
taking the logarithm of the partition function. This provides a further independent check of the result.
The validity of the method of expansion by regions is argued in appendix \ref{app:region},  here we mainly exploit the combinatorics coming from the labeling of the regions.

The simple fact on which Mayer expansion is based on is
\begin{equation}
 \prod_{1\leq I<J\leq k}(1+f_{IJ})\,=\,
\sum_{g\,\in\,\mathcal{G}^{[k]}}\,\prod_{e(I,J)\,\in\,g}\,f_{IJ}\,=\,
\sum_{\mathsf{b}\,\in\,\mathcal{B}^{[k]}}\,\prod_{\mathbf{Y}\,\in\,\mathsf{b}}\,
\left(\sum_{g\,\in\,\mathcal{G}_c^\mathbf{Y}}\prod_{e(I,J)\,\in\,g}\,f_{IJ}\right)\,.
\label{simpleMayer}
\end{equation}
For convenience we summarize the notation used in the text here: $\mathcal{G}^{\mathbf{X}}$ (respectively $\mathcal{G}_c^{\mathbf{X}}$), where $\mathbf{X}$ is a set, denotes the set of graphs (respectively connected graphs)
with the elements of $\mathbf{X}$ as vertices. $\mathcal{B}^{\mathbf{X}}$ is the set of grouping of the elements of   $\mathbf{X}$ into clusters.
We set $[k]=\{1,2,\dots,k\}$. Finally, $e(I,J)\,\in\,g$ means that there is an edge connecting the vertices $I$ and $J$ in the graph $g$.

The rewriting \eqref{simpleMayer} together with basic properties of the exponential implies that only connected graphs contibutes to the logarithm of the grand canonical partition function.
The expansion \eqref{simpleMayer} is not particularly convenient when expanding the integrand in a region, see Section \ref{byregions}.
There is a refined version of \eqref{simpleMayer}  that we will now present. 
For any fixed $\mathsf{b}\,\in\,\mathcal{B}^{[k]}$ we can rearrange the integrand as
\begin{equation}
\prod_{I=1}^k\,U_I  \prod_{1\leq I<J\leq k}(1+f_{IJ})\,=\,
\prod_{\mathbf{Y}\,\in\,\mathsf{b}}\,\mathfrak{U}_{\mathbf{Y}}\,
\prod_{\substack{\mathbf{Y}_1,\mathbf{Y}_2\,\in\,\mathsf{b} \\ \mathbf{Y}_1<\mathbf{Y}_2}}\left(1+\mathfrak{f}_{\mathbf{Y}_1\mathbf{Y}_2}\right)\,,
\label{cluster2}
\end{equation}
where
\begin{equation}
 \mathfrak{U}_{\mathbf{Y}}\,:=\,\prod_{I\,\in\,\mathbf{Y}}\,U_I\,
\prod_{\substack{I,J\,\in\,\mathbf{Y} \\ I<J}}(1+f_{IJ})\,,
\qquad \qquad 
1+\mathfrak{f}_{\mathbf{Y_1}\mathbf{Y_2}}\,:=\,\prod_{\substack{I_1\,\in\,\mathbf{Y}_1 \\ I_2\,\in\,\mathbf{Y}_2}}\,(1+f_{I_1 I_2})\,.
\end{equation}
The factors $ \mathfrak{U}_{\mathbf{Y}}$,  $ \mathfrak{f}_{\mathbf{Y}_1,\mathbf{Y}_2}$ respectively takes into account ``interactions'' within each cluster $\mathbf{Y}$ and  between
 pairs of clusters $\mathbf{Y}_1$, $\mathbf{Y_2}$. 
Applying the same reasoning as in  \eqref{simpleMayer} to the right hand side of  \eqref{cluster2} we rewrite
\begin{equation}
 \prod_{I=1}^k\,U_I \prod_{1\leq I<J\leq k}(1+f_{IJ})\,=\,
\prod_{\mathbf{Y}\,\in\,\mathsf{b}}\,\mathfrak{U}_{\mathbf{Y}}\,
\sum_{\mathsf{b}'\,\in\,\mathcal{B}^{[N_{\mathsf{b}}]}}\,\prod_{\mathbf{Y}'\,\in\,\mathsf{b}'}\,
\left(\sum_{g\,\in\,\mathcal{G}_c^{\mathbf{Y}'}}\prod_{e(\mathbf{Y}_1,\mathbf{Y}_2)\,\in\,g}\,\mathfrak{f}_{\mathbf{Y_1}\mathbf{Y_2}}\right)\,.
\label{cluster3}
\end{equation}
 The number $N_{\mathsf{b}}$ denotes the number of clusters in the grouping $\mathsf{b}$.
The  sum over $\mathcal{B}^{[N_{\mathsf{b}}]}$  in \eqref{cluster3}  is interpreted as  a sum over clusters of clusters.

Assuming the validity of the method of expansion by regions we have
\begin{equation}
\mathcal{Z}\,=\,\sum_{k=0}^{\infty}\,\frac{q^k}{k!}\,\sum_{\mathsf{b}\,\in\,\mathcal{B}^{[k]}}\,\int_k\,\mathbb{T}^{(\mathsf{b})}\left[\text{Integrand}_k\right]\,,
\end{equation}
where we define $\text{Integrand}_k$ as the r.h.s.~of \eqref{cluster3}.
The symbol $\mathbb{T}^{(\mathsf{b})}\left[\dots\right]$ means the following: Taylor expand its argument in the region labeled by $\mathsf{b}$.
Based  purely on combinatorics  one rewrites
\begin{equation}
\log\,\mathcal{Z}\,=\,\sum_{k=0}^{\infty}\,\frac{q^k}{k!}\,\sum_{\mathsf{b}\,\in\,\mathcal{B}^{[k]}}\,
\int_k\,\mathbb{T}^{(\mathsf{b})}
\left[
\prod_{\mathbf{Y}\,\in\,\mathsf{b}}\mathfrak{U}_{\mathbf{Y}}\,
\sum_{g\,\in\,\mathcal{G}_c^{[N_{\mathsf{b}}]}}\,\prod_{e(\mathbf{Y}_1,\mathbf{Y}_2)\,\in\,g}\,\mathfrak{f}_{\mathbf{Y}_1,\mathbf{Y}_2}\right]\,.
\label{FullLog}
\end{equation}
The explicit Taylor expansion gives
\begin{equation}
 \mathbb{T}^{(\mathsf{b})}\left[\mathfrak{U}_{\mathbf{Y}}\right]\,=\,
\left(\frac{Q(\phi_{\mathbf{Y}})}{\epsilon_2}\right)^{|\mathbf{Y}|}\,
\prod_{\stackrel{I,J\,\in\,\mathbf{Y}}{I<J}}\,
\frac{\phi_{IJ}^2}{\phi_{IJ}^2-\epsilon_2^2}+\dots\,,
\qquad \qquad 
\phi_{\mathbf{Y}}:=\frac{1}{|\mathbf{Y}|}\,\sum_{I\,\in\,\mathbf{Y}}\,\phi_I\,,
\label{UUdef}
\end{equation}
\begin{equation}
  \mathbb{T}^{(\mathsf{b})}\left[\mathfrak{f}_{\mathbf{Y}_1,\mathbf{Y}_2}\right]\,=\,
\epsilon_2\,|\mathbf{Y}_1|\,|\mathbf{Y}_2|\,G(\phi_{\mathbf{Y}_1}-\phi_{\mathbf{Y}_2})+\dots
\label{getPROPA}
\end{equation}
From \eqref{getPROPA} one concludes immediatly that only connected \emph{tree graphs} contributes to the sum in \eqref{FullLog}.
This produces a factor of $\epsilon_2^{N_{\mathsf{b}}-1}$. 
Next, for each cluster we separate the integration variables into center of the cluster and distances between particles in the same cluster.
The latter contribution facorizes from  \eqref{UUdef} to produce
\begin{equation}
\frac{1}{\epsilon_2^{|\mathbf{Y}|}}\,\int\,\prod_{I\,\in\,\mathbf{Y}}
\frac{d\phi_I}{2\pi i}\,\delta\left( \frac{1}{|\mathbf{Y}|}\sum_{I\,\in\,\mathbf{Y}}\phi_I\right)\,
\prod_{\stackrel{I,J\,\in\,\mathbf{Y}}{I<J}}\,
\frac{\phi_{IJ}^2}{\phi_{IJ}^2-\epsilon_2^2}\,=\,\frac{1}{\epsilon_2}\,\frac{1}{2\pi i}\frac{|\mathbf{Y}|!}{|\mathbf{Y}|^2}\,.
\end{equation}
We have thus reproduced the result \eqref{resultMayer2}.



\end{document}